\documentclass[12pt]{article}
\usepackage{amsmath,amsthm,amssymb}
\usepackage{graphicx,epstopdf,framed,color}
\usepackage{euscript,paralist,wrapfig}

\theoremstyle{remark}
\newtheorem{theorem}{Theorem}[section]
\newtheorem{corollary}[theorem]{Corollary}
\newtheorem{lemma}[theorem]{Lemma}

\newtheorem{proposition}[theorem]{Proposition}

\newtheorem{remark}[theorem]{Remark}

\newtheorem{problem}[theorem]{Problem}

\numberwithin{equation}{section}

\newfont{\roma}{cmr10 scaled 1200}
\renewcommand{\cline}{{\mathbb C}}

\newcommand{\nline}  {{\mathbb N}}
\newcommand{\rline}  {{\mathbb R}}

\newcommand{\tline}  {{\mathbb T}}

\newcommand{\dd}   {{\rm d}\hbox{\hskip 0.5pt}}
\newcommand{\DD}   {{\rm D}\hbox{\hskip 0.5pt}}
\newcommand{\Ascr} {{\cal A}}

\newcommand{\Dscr} {{\cal D}}

\newcommand{\Iscr} {{\cal I}}
\newcommand{\Jscr} {{\cal J}}

\newcommand{\Oscr} {{\cal O}}

\newcommand{\mm}    {{\hbox{\hskip 0.5pt}}}
\newcommand{\m}     {{\hbox{\hskip 1pt}}}

\newcommand{\bluff} {{\hbox{\raise 15pt \hbox{\mm}}}}
\newcommand{\sbluff}{{\hbox{\raise  7pt \hbox{\mm}}}}

\newcommand{\FORALL} {{\hbox{$\hskip 11mm \forall \;$}}}

\renewcommand{\Re}   {{\rm Re\,}}

\newcommand{\bbm}[1]{\left[\begin{matrix} #1 \end{matrix}\right]}
\newcommand{\sbm}[1]{\left[\begin{smallmatrix} #1
   \end{smallmatrix}\right]}

\begin{document}
\renewcommand{\thefootnote}{\fnsymbol{footnote}}
\renewcommand{\thefootnote}{\fnsymbol{footnote}}
\newcommand{\footremember}[2]{%
   \footnote{#2}
    \newcounter{#1}
    \setcounter{#1}{\value{footnote}}%
}
\newcommand{\footrecall}[1]{%
    \footnotemark[\value{#1}]%
}
\makeatletter
\def\blfootnote{\gdef\@thefnmark{}\@footnotetext}
\makeatother

\begin{center}
{\Large \bf Motion planning for parabolic equations using \\[1ex]
  flatness and finite-difference approximations}\\[2ex]
Soham Chatterjee and Vivek Natarajan \vspace{-2mm}
\blfootnote{This work was supported by the Science and Engineering Research Board, DST India, via the grant ECR/2017/002583.}
\blfootnote{S. Chatterjee (soham.chatterjee@iitb.ac.in) and V. Natarajan (vivek.natarajan@iitb.ac.in) are with the Systems and Control Engineering Group, Indian Institute of Technology Bombay, Mumbai, India, 400076, Ph:+912225765385.}
\end{center}

\begin{abstract} {{\noindent
We consider the problem of finding an input signal which transfers a linear boundary controlled 1D parabolic partial differential equation with spatially-varying coefficients from a given initial state to a desired final state. The initial and final states have certain smoothness and the transfer must occur over a given time interval. We address this motion planning problem by first discretizing the spatial derivatives in the parabolic equation using the finite-difference approximation to obtain a linear ODE in time. Then using the flatness approach we construct an input signal that transfers this ODE between states determined by the initial and final states of the parabolic equation. We prove that, as the discretization step size converges to zero, this input signal converges to a limiting input signal which can perform the desired transfer for the parabolic equation. While earlier works have applied this motion planning approach to constant coefficient parabolic equations, this is the first work to investigate and establish the efficacy of this approach for parabolic equations with discontinuous spatially-varying coefficients. Using this approach we can construct input signals which transfer the parabolic equation from one steady-state to another. We show that this approach yields a new proof for the null controllability of 1D linear parabolic equations containing discontinuous coefficients and also present a numerical scheme for constructing a null control input signal when the initial state is piecewise continuous. \vspace{-2mm} }}
\end{abstract}

 \vspace{-3mm}

\section{Introduction} \label{sec1} \vspace{-1mm}
\setcounter{equation}{0} 

\ \ \ Transferring boundary controlled partial differential equations (PDEs) between two given states, especially steady-states, over a finite time interval is a problem of practical significance. This problem is central to applications such as start-up and shutdown of fixed-bed tubular reactors, set point change in industrial glass feeders, control of open channel flows, adaptive optics and reorientation of flexible manipulators, see \cite{FlMoRoRu:1998}, \cite{MaSa:2012}, \cite{RaMe:2010}, \cite{RuDoRoOsSa:2013}, \cite{ScMeKu:2013}. A well-known approach to this motion-planning problem is using flatness. The main idea of flatness is to express the state and the input of the differential equation as functions of a flat output and its time derivatives. Then, based on the desired motion, an appropriate trajectory is selected for the flat output using which the input necessary to execute the motion is computed. In this work, we address the motion planning problem for linear boundary controlled 1D parabolic PDEs with spatially-varying coefficients using flatness and finite-difference approximations.

In the predominant approach to the flatness-based motion planning for parabolic PDEs, the PDE state is expressed as an infinite series using a flat output and certain generating functions (which are functions of the spatial variable), and this series determines the series for the PDE input. Using $\{1,x, x^2,\ldots\}$ as generating functions, this approach has been used to solve the motion planning problem for linear and quasilinear 1D heat equations with analytic coefficients in \cite{LaMaRo:2000} and \cite{LyRu:2002}, respectively. Building on these results, a new proof of null controllability for a linear nD constant coefficient heat equation was developed in \cite{MaRoRo:2014}. In \cite{LaMa:2000}, using generating functions obtained by solving a sequence of two-point boundary-value problems, the motion planning problem was solved for linear 1D heat equation with analytic coefficients. More recently in \cite{MaRoRo:2016}, using generating functions obtained by solving a sequence of Cauchy problems, the motion planning problem was solved for linear 1D heat equations with $L^p$ coefficients to establish their null controllability.

Flatness-based motion planning of linear parabolic PDEs with smooth coefficients defined on higher dimensional parallelepipedons has been addressed in \cite{Meu:2011} and \cite{MeKu:2009}. In \cite{Meu:2011}, taking the Laplace transform of the spectral representation of the PDE, the PDE state and input are expressed as functions of a flat output. In \cite{MeKu:2009} the PDE is integrated formally to obtain an implicit parametrization of the PDE state and input in terms of a flat output via a Volterra–type integral equation. This idea has been generalized in \cite{ScMeJu:2013} to consider systems of 1D semilinear parabolic PDEs with smooth coefficients.

A natural approach to the motion planning of PDEs is as follows: (i) discretize the spatial derivative in the PDE using the finite-difference scheme to obtain an $n^{\rm th}$-order ODE in time, (ii) find an input $f_n$, using the flatness technique, which solves an appropriate motion planning problem for the ODE and (iii) prove that as the discretization step size tends to zero, i.e. as $n\to\infty$, this $f_n$ converges to a limiting input which can transfer the PDE between the given initial and final states. In \cite{OlSe:2001} this approach was used to address the motion planning problem for 1D constant coefficient heat equations with only the diffusion term. In \cite{UtMeKu:2010} this approach was proposed to transfer 1D quasilinear parabolic PDEs, whose coefficients are analytic functions of the state, between steady-states. It was shown that the proposed approach works, under a certain analyticity assumption, when the coefficients are some affine functions of the state.

The motion planning approach for PDEs via semi-discretization and flatness (discussed above) provides a simple numerical scheme, based on the application of the flatness technique to ODEs, for computing the desired input for the PDE. This approach can be applied, in principle, to any 1D boundary controlled PDE. Indeed, it has been applied (at times without proof) to a flexible rod model in \cite{OlSe:2001}, Euler-Bernoulli beam in \cite{ChNa:2020}, a first-order PDE in \cite{RiSi:2018} and some parabolic equations (as mentioned above). For these reasons, this motion planning approach is attractive from an application standpoint. However, papers on this approach are few and have restrictive assumptions. In particular, existing works applying this approach to parabolic PDEs require the coefficients to be infinitely smooth so that the solution of the PDE can be written as a power series. We overcome this limitation in this paper. We prove that the motion planning approach for PDEs via semi-discretization and flatness can be applied to linear 1D parabolic PDEs whose coefficients are discontinuous to transfer the PDE between certain smooth states, in particular steady states. Our results yield a new proof for the null controllability of linear 1D parabolic PDEs and a numerical scheme for constructing a null control input signal when the initial state is piecewise continuous. A preliminary version of some results in this paper are in our conference paper \cite{ChNa:2020}.

The paper is organized as follows. We describe a motion planning problem for a 1D parabolic PDE in Section \ref{sec2} and present a solution to this problem (sans proof) in Section \ref{sec2.1}. In Section \ref{sec3} we prove that the solution of a semi-discrete approximation of the 1D parabolic PDE converges to the solution of the PDE as the discretization step size is reduced to zero. Section \ref{sec4} compares two choices for the initial and final states of the semi-discrete approximation of the PDE. Section \ref{sec5} contains the proof of our solution (in Section \ref{sec2.1}) to the motion planning problem. In Section \ref{sec6}, we present our result on the null controllability of the 1D parabolic PDE. We illustrate our theoretical results numerically in Section \ref{sec7}.


{\em Notations and definitions}: Let $C^{k}[0,1]$ be the space of $k$-times continuously differentiable functions on $[0,1]$ with the standard norm. A function $\psi$ on $[0,1]$ is in piecewise $C^k[0,1]$, written as $PC^k[0,1]$, if the following holds: there exists a finite partitioning of $[0,1]$ into disjoint intervals such that on each of the these intervals $\psi$ is the restriction of some $C^k[0,1]$ function to that interval. For $\psi\in PC^k[0,1]$, we let $\|\psi\|_{PC^k[0,1]}$ to be the supremum of $\|\psi\|_{C^k[a,b]}$ over all the intervals $[a,b]\subset[0,1]$ on which $\psi$ is $k$-times continuously differentiable. We write $PC^0[0,1]$ as $PC[0,1]$. Let $H^2(0,1)$ be the usual Sobolev space of order $2$. A function $\psi\in C^1[0,1]\cap H^2(0,1)$ is in $PC^{(2),1}[0,1]$ if $\psi_{xx}\in PC^1[0,1]$. For $\psi\in PC^{(2),1}[0,1]$, we let $\|\psi\|_{PC^{(2),1}[0,1]}= \|\psi\|_{C^1[0,1]}+\|\psi_{xx}\|_{PC^1[0,1]}$. For functions on $[0,1]$, we often omit $[0,1]$ in the notation of their norms. For a smooth function $y:[0,T]\to\rline$, let $y^{(m)}$ denote its $m^{\rm th}$-derivative. A smooth function $y:[0,T] \to \rline$ is said to be of Gevrey order $\alpha>0$ if there exists a positive constant $D$ such that $\sup_{t\in[0,T]}|y^{(m)}(t)|\leq D^{m+1}(m!)^{\alpha}$ for all $m\in \nline$. We denote the class of all functions satisfying these estimates by $G_\alpha[0,T]$.

For $v= [v_1 \ v_2 \ \cdots \ v_n]^\top\in\rline^n$, let $\|v\|_p=\big(\sum_{i=1}^n |v_i|^p\big)^\frac{1}{p}$ for $p\geq1$,
$\|v\|_\infty= \max_{1\leq j \leq n} |v_j|$ and $\|v\|_{2d}=\sqrt{h}\, \|v\|_2$, where $h=1/(n+1)$. The extension operator $S_n:\rline^n \to L^\infty(0,1)$ is defined as follows: $[S_n v](0)=v_1$, $[S_n v](x)=v_j$ for $j\in \{1,2,\ldots n\}$ and $(j-1)h < x \leq j h$ and $[S_n v](x)=0$ for $nh < x\leq 1$. For a function $z\in PC[0,1]$, we let $R_n z = [z(h) \  z(2h) \ \cdots \ z(nh)]^\top$. We will take any summation $\sum_{i=j}^k p_i$ to mean zero whenever $k<j$. 

\section{Problem and solution} \label{sec2} \vspace{-1mm}
\setcounter{equation}{0} 

Consider the following linear 1D parabolic PDE on the interval $x\in[0,1]$ and $t\geq0$:
\begin{align}
 & u_t(x,t) = \theta(x) u_{xx}(x,t) + \sigma(x) u_x(x,t) + \lambda(x)u(x,t), \label{eq:heat1} \\[0.5ex]
 & \alpha_0 u_x(0,t) + \beta_0 u(0,t)=0, \qquad \alpha_1 u_x(1,t) + \beta_1 u(1,t)= f(t), \label{eq:heat2}
\end{align}
where $f$ is the input. The coefficients $\theta$, $\sigma$ and $\lambda$ are in $PC^1[0,1]$ and $\inf_{x\in[0,1]}\theta(x)>0$.
Let $\Iscr$ be the collection of all the points where either $\theta$, $\sigma$ or $\lambda$ is not differentiable (so $\Iscr$ contains the points where these functions are discontinuous). The PDE \eqref{eq:heat1}-\eqref{eq:heat2} with $f=0$ can be written as an evolution equation on $L^2(0,1)$ as follows: $u_t(\cdot,t) = \Ascr u(\cdot,t)$, where the domain of $\Ascr$ is $ \Dscr(\Ascr) = \{ w \in H^2(0,1) \m\big|\m \alpha_0 w_x(0) + \beta_0 w(0) = \alpha_1 w_x(1) + \beta_1 w(1)=0 \}$
and for all $w\in \Dscr(\Ascr)$ and $x\in(0,1)$,
$$ \Ascr w(x) = \theta(x) w_{xx}(x) + \sigma(x) w_x(x) + \lambda(x) w(x). $$
The graph norm on $\Dscr(\Ascr)$ is $\|w\|_{\Dscr(\Ascr)}=\|\Ascr w\|_{L^2}+\|w\|_{L^2}$ for $w\in\Dscr(\Ascr)$. We will let $\Ascr w$ be the function obtained by applying $\Ascr$ (as a differential operator) to $w\in H^2(0,1)$ even if $w\notin \Dscr(\Ascr)$. From the spectral result \cite[Proposition 2.2]{MaRoRo:2016} it follows that $\Ascr_0$ defined on $\Dscr(\Ascr)$ as $\Ascr_0 w=\theta w_{xx}$ generates an analytic semigroup on $L^2(0,1)$. Using this, the norm estimate
$\|\Ascr w - \Ascr_0 w\|_{L^2} \leq \epsilon  \|\Ascr_0 w\|_{L^2} + N(\epsilon) \|w\|_{L^2}$ for each $\epsilon>0$ and some $N(\epsilon)>0$ which can be inferred from \cite[Chapter III,  Example 2.2.]{EnNa:2006} and the perturbation result \cite[Chapter 3, Theorem 2.1]{Pazy:1983}, it follows that $\Ascr$ generates an analytic semigroup on $L^2(0,1)$.
The above norm estimate implies that $\|w\|_{H^2(0,1)} \leq c  \|w\|_{\Dscr(\Ascr)}$ for all $w\in\Dscr(\Ascr)$ and some $c>0$.
We will often write $u(\cdot,t)$ as $u(t)$. When $f\in C^3[0,T]$ and the initial state $u_0\in L^2(0,1)$, the state trajectory $u$ for \eqref{eq:heat1}-\eqref{eq:heat2} on the time interval $[0,T]$ is the continuous $L^2(0,1)$-valued function given by \vspace{-2mm}
\begin{equation} \label{eq:mildsoln}
 u(t) = \tline_t \left[u_0 - \nu f(0) \right] + \int_0^t \tline_{t-\tau}\big[ \Ascr \nu f(\tau)- \nu \dot{f}(\tau)\big]\dd \tau + \nu f(t). \vspace{-1mm}
\end{equation}
Here $\nu(x)= \mu_1 x^{\mu_2}$ for $x\in[0,1]$ with $\mu_1,\mu_2\in \rline$ chosen so that $\mu_2\geq3$ and $\mu_1(\mu_2 \alpha_1+\beta_1)=1$. The above expression for $u$ is obtained by transforming \eqref{eq:heat1}-\eqref{eq:heat2} into another PDE \eqref{eq:wheat1}-\eqref{eq:wheat2} whose boundary conditions are homogeneous, see Lemma \ref{lm:reg}. In this paper we address the following motion planning problem. \vspace{-1mm}

\begin{framed} \vspace{-2mm}
\begin{problem} \label{prob:motionplan}
Let a time $T>0$, an initial state $u_0\in PC^{(2),1}[0,1]$ and a final state $u_T\in PC^{(2),1}[0,1]$ satisfying $\alpha_0 u_{0,x}(0)+\beta_0 u_0(0)=0$ and $\alpha_0 u_{T,x}(0)+\beta_0 u_T(0)=0$ be given. Suppose that $\Ascr^k u_0$ and $\Ascr^k u_T$ belong to $\Dscr(\Ascr)$ and satisfy the estimates
\begin{equation} \label{eq:probest}
 \|\Ascr^k u_0\|_{L^2} \leq (C_0)^k\m k! \, , \qquad \|\Ascr^k u_T\|_{L^2} \leq (C_T)^k\m k!  \FORALL k\in\{1,2,\ldots\}
\end{equation}
and some constants $C_0,C_T>0$. Find $f\in C^\infty[0,T]$ such that the state trajectory $u$ of \eqref{eq:heat1}-\eqref{eq:heat2} with initial state $u_0$ and input $f$ satisfies \vspace{-2mm} $u(T)=u_T$.
\end{problem}
\end{framed}
\vspace{-2.5mm}
In the above problem, we do not need the functions $u_0$ and $u_T$ to be in $\Dscr(\Ascr)$, but require that $\Ascr u_0$ and $\Ascr u_T$ (obtained by applying $\Ascr$ as a differential operator to $u_0$ and $u_T$) belong to $\Dscr(\Ascr^\infty)$. An example of such functions which are in general not in $\Dscr(\Ascr)$ are steady states of \eqref{eq:heat1}-\eqref{eq:heat2}, see Remark \ref{rm:ss}. The hypothesis on $u_0$ and $u_T$ in Problem \ref{prob:motionplan} are also satisfied by the function $\tline_s z$ for any $s>0$ and $z\in L^2(0,1)$  and $\tline_s z$ is in $\Dscr(\Ascr)$, see beginning of Section \ref{sec6} for details. A solution to Problem \ref{prob:motionplan} can be used to transfer the PDE \eqref{eq:heat1}-\eqref{eq:heat2} between steady states, see Remark \ref{rm:ss}, and to establish its null controllability, see Section \ref{sec6}. \vspace{-4mm}

\subsection{Solution to the motion planning problem} \label{sec2.1}
\vspace{-1mm}

\ \ \ In this section, we describe our solution to Problem \ref{prob:motionplan} (detailed proofs are in Sections \ref{sec3}, \ref{sec4} and \ref{sec5}). The $n^{\rm th}$-order semi-discrete approximation of \eqref{eq:heat1}-\eqref{eq:heat2} with input $f=f_n$, obtained via finite-difference discretization of the spatial derivatives, is \vspace{0mm}
\begin{equation} \label{eq:semi-disc}
 \dot v_n(t) = A_n v_n(t) + B_nf_n(t) \FORALL t\geq0. \vspace{0mm}
\end{equation}
Here $v_n(t)=[v_{n,1}(t) \ \ v_{n,2}(t) \ \cdots \ v_{n,n}(t)]^\top \in \rline^n$ and the matrices $A_n \in \rline^{n\times n}$ and $B_n \in \rline^{n\times 1}$ are defined as follows: $A_n = \Theta_n L_n + \Sigma_n D_n + \Lambda_n$ with $\Theta_n, L_n,\Sigma_n, D_n,\Lambda_n\in\rline^{n\times n}$,
$\Theta_n = \text{diag}\bbm{\theta(h) & \theta(2h) & \cdots & \theta(nh)}$, $\Sigma_n = \text{diag}\bbm{\sigma(h) & \sigma(2h) & \cdots & \sigma(nh)},$
$\Lambda_n = \text{diag}\bbm{\lambda(h) & \lambda(2h) & \cdots & \lambda(nh)},$ \vspace{-2mm}
$$ L_n = \frac{1}{h^2}
      \bbm{
        -2+4 r_0 & 1-r_0 & 0 & \cdots & \cdots &\cdots &0 \\
        1 & -2 & 1 & 0 &\cdots &\cdots &0 \\
        0 &  1 &-2 & 1 &0 &\cdots &0  \\
        \vdots  & \ddots & \ddots & \ddots & \ddots & \ddots & \vdots \\
        0 & \cdots & 0 &1 & -2 & 1 &0 \\
        0 & \cdots & \cdots &0 & 1  &-2 &1\\
        0 & \cdots & \cdots &\cdots &0 &1-r_1 &-2+4 r_1}, \vspace{-1mm} $$
$$ D_n = \frac{1}{h}
    \bbm{
      h q_0& 0 & \cdots & \cdots &\cdots &0 \\
      -1 & 1 & 0 &\cdots &\cdots &0 \\
      0 &  -1 &1 &0 &\cdots &0  \\
      \vdots  & \ddots & \ddots & \ddots & \ddots & \vdots \\
      0 & \cdots  &0 & -1 & 1 &0 \\
      0 & \cdots & \cdots &0  &-1 &1},    \quad
  B_n = \frac{1}{h^2}\bbm{0 \\ 0 \\ \vdots \\ 0 \\ 0 \\ b_n}. \vspace{-1mm} $$
Here $h=1/(n+1)$, $r_0=\alpha_0/(3\alpha_0-2h\beta_0)$, $r_1=\alpha_1/(3\alpha_1+2h\beta_1)$, $q_0 = -\beta_0/(\alpha_0-h\beta_0)$ and $b_n = 2h\theta(nh)/(3\alpha_1+2h\beta_1)$. While $r_0$, $r_1$, $q_0$ and $b_n$ are well-defined for all $n\gg1$, i.e. for all $n$ sufficiently large, to simplify our presentation we assume that they are well-defined for all $n$. This is reasonable since we are interested in the solutions of \eqref{eq:semi-disc} in the limit $n\to\infty$.
By rearranging the terms, the system of equations \eqref{eq:semi-disc} can be written equivalently as \vspace{0mm}
\begin{equation}
  v_{n,2} = \frac{h^2}{(1- r_0)\theta(h)} \dot v_{n,1} - \frac{h^2 \big[ \sigma(h) q_0 + \lambda(h) \big]+ (4r_0-2)\theta(h)} {(1-r_0)\theta(h)}v_{n,1}, \label{eq:vn2flat} \vspace{0mm}
\end{equation}
\begin{align}
  v_{n,j+1} &= \frac{h^2}{\theta(jh)} \dot{v}_{n,j} + \frac{h\sigma(jh)-\theta(jh)}{\theta(jh)} v_{n,j-1} + \frac{2\theta(jh)-h\sigma(jh)-h^2\lambda(jh)}{\theta(jh)}v_{n,j} \nonumber\\
  &\hspace{40mm} \FORALL j\in\{2,3,\ldots n-1\}, \label{eq:vnjflat}\\[-6.5ex]\nonumber
\end{align}
\begin{align}
  f_n &= \frac{h^2}{b_n} \dot{v}_{n,n} - \frac{(1-r_1)\theta(nh)-h\sigma(nh)}{b_n} v_{n,n-1}  \nonumber\\
  &\hspace{20mm} - \frac{(4r_1-2)\theta(nh)+h\sigma(nh)+\lambda(nh) h^2}{b_n} v_{n,n}  \label{eq:fnflat}.\\[-4.7ex]\nonumber
\end{align}
From the above expressions, it follows that $v_{n,j}$ for $j\in \{2,3,\ldots,n\}$ and $f_n$ can be expressed as linear combinations of $v_{n,1}$ and its derivatives $v^{(1)}_{n,1}, v^{(2)}_{n,1},\ldots v^{(n)}_{n,1}$. So \eqref{eq:semi-disc} is a flat system with the flat output \vspace{-2mm}
\begin{equation} \label{eq:yn}
 y_n(t) =  (\alpha_0 - q_0\beta_0) v_{n,1}(t) \vspace{-2mm}
\end{equation}
and, using the above expressions, we can compute constants $a_{n,m}$ such that \vspace{-3.5mm}
\begin{equation}\label{eq:param}
  f_n(t) = \sum_{m=0}^{n} a_{n,m} y_n^{(m)}(t). \vspace{-3.5mm}
\end{equation}
Similar definitions of the flat output and state and input parameterizations using it have appeared in \cite{OlSe:2001}, \cite{UtMeKu:2010}. Naturally, given $y_n\in C^\infty[0,\infty)$, the solution $v_n$ of \eqref{eq:semi-disc} for the input $f_n$ determined by \eqref{eq:param} and the initial condition $v_n(0)$ determined by $y_n(0), y_n^{(1)}(0),\ldots y_n^{(n-1)}(0)$ via \eqref{eq:yn} and \eqref{eq:vn2flat}-\eqref{eq:vnjflat} satisfies \eqref{eq:yn} for all $t\geq0$.

Given $u_0$ and $u_T$ as in Problem \ref{prob:motionplan}, fix $y\in G_\alpha[0,T]$ with $1<\alpha<2$ such that \vspace{-2mm}
\begin{align}
  y^{(m)}(0) &= \alpha_0 \Ascr^m u_0(0) - \beta_0 (\Ascr^m u_0)_x(0) \FORALL m\geq0,\label{eq:y_endpts1}\\
  y^{(m)}(T) &= \alpha_0 \Ascr^m u_T(0) - \beta_0 (\Ascr^m u_T)_x(0) \FORALL m\geq0. \label{eq:y_endpts2} \\[-4.5ex]\nonumber
\end{align}
Such a $y$ can always be found, see Proposition \ref{prop:yconst}, and it is particularly easy to construct when $u_0$ and $u_T$ are steady-states. Compute $f_n\in C^\infty[0,T]$ by letting $y_n = y$ in \eqref{eq:param}. Then there exists $f\in C^\infty[0,T]$ such that \vspace{-2mm}
$$ \lim_{n\to\infty}\|f_n - f\|_{C^1[0,T]} = 0 \vspace{-2mm}$$
and $f$ solves Problem \ref{prob:motionplan} for the given $u_0, u_T$. We establish this result in Theorem \ref{thm:finalmainresult}. Clearly, for $n$ large enough $f_n$ computed in \eqref{eq:param} solves Problem \ref{prob:motionplan} approximately, i.e. $\|u(T)-u_T\|_{L^2}$ will be small for the solution of \eqref{eq:heat1}-\eqref{eq:heat2} with input $f_n$ and initial state $u(0)=u_0$. We remark that to obtain a good estimate for $f_n$, it may be sufficient to use only the first several terms in the summation in \eqref{eq:param}.

\begin{remark} \label{rm:ss}
A steady-state for \eqref{eq:heat1}-\eqref{eq:heat2} is $u_{ss}\in H^2(0,1)$ for which $\Ascr u_{ss}=0$, $\alpha_0 u_{ss,x}(0)+\beta_0 u_{ss}(0)=0$ and $\alpha_1 u_{ss,x}(1)+\beta_1 u_{ss}(1)=f_{ss}$ for some $f_{ss}\in\rline$. Mimicking the steps above \eqref{eq:AkPC2est} with $u_0=u_{ss}$ and $k=0$ we get $u_{ss}\in PC^{(2),1}[0,1]$. The solution to Problem \ref{prob:motionplan} presented in this section can be used to find an input $f\in C^\infty[0,T]$ which transfers \eqref{eq:heat1}-\eqref{eq:heat2} between any two steady states $u_0$ and $u_T$ over the time interval $[0,T]$ since $u_0$ and $u_T$ satisfy the estimates in \eqref{eq:probest} trivially. \vspace{-3mm}
\end{remark}

\section{Finite-difference based semi-discretization} \vspace{-2mm} \label{sec3} \setcounter{equation}{0} 

\ \ \ In this section, we approximate the spatial derivatives in the PDE \eqref{eq:heat1}-\eqref{eq:heat2} using the finite-difference method to obtain an ODE in time, referred to as the semi-discrete approximation of the PDE. We show that the solution to this ODE converges to the solution of the PDE as the discretization step size is reduced to zero. While many works have established similar convergence results when both the time and spatial derivatives in the PDE are approximated using the finite-difference method, see for instance \cite{Joh:1982}, \cite{Jov:1989}, few works have studied the convergence of the semi-discrete approximations. In particular, the convergence results required in this paper (see Theorem \ref{th:semidisc_conv} and Corollary \ref{cor:semidisc_conv}) do not follow from existing works.

Consider a function $u\in C([0,T];PC^{(2),1}[0,1])$ such that $u(\cdot, t)$ satisfies the boundary conditions in \eqref{eq:heat2} for each $t\in[0,T]$. Let the points at which $u_{xx}(\cdot,t)$ is not differentiable be a subset of $\Iscr$ (defined below \eqref{eq:heat2}) for all $t\geq0$. For a positive integer $n$, divide $[0,1]$ into $n+1$ intervals and let $h=1/(n+1)$. Recall $r_0$, $r_1$ and $q_0$ defined above \eqref{eq:vn2flat}. Define the discrete version $\Delta_n u(jh,t)$ of $u_{xx}(jh,t)$ as follows:
\vspace{-0.5mm}
\begin{equation} \label{eq:Deltan1}
 \hspace{-50mm}\Delta_n u(h,t) = \frac{1}{h^2} [(-2+4 r_0) u(h,t)+(1-r_0) u(2h,t)], \vspace{-0.5mm}
\end{equation}
\begin{equation} \label{eq:Deltanj}
 \Delta_n u(jh,t) = \frac{1}{h^2} [u(jh-h,t)-2u(jh,t)+u(jh+h,t)] \quad\forall\, j\in \{2,3,\ldots n-1\}, \vspace{-0.5mm}
\end{equation}
\begin{equation} \label{eq:Deltann}
 \Delta_n u(nh,t) = \frac{1}{h^2} \left[(1-r_1)u(nh-h,t) - (2-4r_1)u(nh,t) + \frac{2hf(t)}{3\alpha_1+ 2h\beta_1}\right]. \vspace{-0.5mm}
\end{equation}
These expressions are obtained via the central-difference approximation of $u_{xx}(jh,t)$ (and incorporating the boundary conditions when $j=1$ and $j=n$). Let $\Jscr_n\subset \{2,3,\ldots n-1\}$ be the set containing every $j$ such that  $\Iscr\cap[jh-h,jh+h]$ is non-empty. The cardinality of $\Jscr_n$ is less than thrice the cardinality of $\Iscr$. Let $\Jscr_n^c=\{1,2,\ldots n\}\setminus \Jscr_n$.
Let $\delta_j=1$ for $j\in \Jscr_n^c$ and $\delta_j=0$ for $j\in \Jscr_n$. For all $n\gg1$, using the Taylor's theorem with the integral form of the remainder (written with $u_{xx}$ as the integrand) we get
\vspace{-2mm}
\begin{equation} \label{eq:appr_uxxC}
 u_{xx}(jh,t) = \Delta_n u(jh,t) + C_{\Delta u}(jh,t) h^{\delta_j}, \vspace{-1mm}
\end{equation}
where $C_{\Delta u}(jh,t)$ satisfies the bound $|C_{\Delta u}(jh,t)| \leq K\sup_{t\in[0,T]} \|u_{xx}(\cdot,t)\|_{PC^1}$ for some constant $K>0$ independent of $u$ and $n$. Define the discrete version $\nabla_n u(jh,t)$ of $u_x(jh,t)$ as follows: \vspace{-1mm}
\begin{align}
 \nabla_n u(h,t) &= q_0 u(h,t), \label{eq:nablan1} \\
 \nabla_n u(jh,t) &= \frac{u(jh,t)-u(jh-h,t)}{h} \FORALL j\in
 \{2,3,\ldots n\}. \label{eq:nablanj}\\[-4.6ex]\nonumber
\end{align}
These expressions are obtained using the backward difference approximation of $u_x(jh,t)$ and incorporating the boundary conditions when $j=1$. Again using the Taylor's theorem with the integral form of the remainder we get \vspace{-0.5mm}
\begin{equation} \label{eq:appr_ux}
  u_x(jh,t) =  \nabla_n u(jh,t) + C_{\nabla u}(jh,t)h, \vspace{-0.5mm}
\end{equation}
where $C_{\nabla u}(jh,t)$ satisfies the bound $|C_{\nabla u}(jh,t)| \leq K\sup_{t\in[0,T]} \|u_{xx}(\cdot,t)\|_{PC[0,1]}$ for some constant $K>0$ independent of $u$ and $n$.

Suppose that the above function $u$ solves \eqref{eq:heat1}-\eqref{eq:heat2} pointwise for $x\in\{h, 2h,\ldots nh\}$ and $t\in[0,T]$. Let $v_n(t)=[u(h,t) \ \ u(2h,t) \ \cdots \ u(nh,t)]^\top$. Then from \eqref{eq:appr_uxxC}-\eqref{eq:appr_ux} we get
\begin{equation} \label{eq:disc}
 \dot v_n(t) = A_n v_n (t) + B_n f(t) + \Oscr_n(t) \FORALL t\in[0,T].
\end{equation}
Here $A_n \in \rline^{n\times n}$ and $B_n \in \rline^{n\times 1}$ are as in \eqref{eq:semi-disc} and $\Oscr_n(t)$ contains the correction terms $C_{\Delta u}(jh,t)h^{\delta_j}$ and $C_{\nabla u}(jh,t)h$. Dropping $\Oscr_n$ from \eqref{eq:disc}, we get the $n^{\rm th}$-order semi-discrete approximation \eqref{eq:semi-disc} of \eqref{eq:heat1}-\eqref{eq:heat2}. In Section \ref{sec3.2} we show that the solution of the $n^{\rm th}$-order ODE \eqref{eq:semi-disc} converges to the solution of the PDE \eqref{eq:heat1}-\eqref{eq:heat2} as $n\to\infty$. \vspace{-1mm}

\subsection{Preliminary lemmas} \label{sec3.1}

\ \ \ We present two simple lemmas in this section. Lemma \ref{lm:reg} discusses the regularity of solutions to \eqref{eq:heat1}-\eqref{eq:heat2}, while Lemma \ref{lm:approxAq} is about the accuracy of the semi-discretization. \vspace{-1mm}

\begin{framed} \vspace{-2mm}
\begin{lemma}\label{lm:reg}
Suppose that $u_0\in L^2(0,1)$ and $f\in C^3[0,T]$. Then $u\in C([0,T];L^2(0,1))$ defined in \eqref{eq:mildsoln} is the unique function in $C((0,T];PC^{(2),1}[0,1])\cap C^1((0,T];C^1[0,1])$ which satisfies \eqref{eq:heat1}-\eqref{eq:heat2} for $t>0$. \vspace{-2mm}
\end{lemma}
\end{framed}
\vspace{-5mm}
\begin{proof}
Recall $\Ascr$ defined below \eqref{eq:heat2} and $\nu$ defined below \eqref{eq:mildsoln}. Although $\nu\notin \Dscr(\Ascr)$, we will write $\Ascr \nu$ to denote $\theta \nu_{xx} + \sigma \nu_x + \lambda \nu$. A formal calculation using \eqref{eq:heat1}-\eqref{eq:heat2} gives that  $w(x,t)= u(x,t) - \nu(x) f(t)$ \vspace{-1mm} satisfies
\begin{align}
 & w_t(t) = \theta w_{xx}(t) + \sigma w_x(t) + \lambda w(t) + \Ascr \nu f(t) - \nu \dot f(t), \label{eq:wheat1} \\
 & \alpha_0 w_x(0,t) + \beta_0 w(0,t)=0, \qquad \alpha_1 w_x(1,t) + \beta_1 w(1,t)= 0. \label{eq:wheat2}\\[-4.3ex]\nonumber
\end{align}
Here we have written $w(\cdot,t)$ as $w(t)$. The mild solution of \eqref{eq:wheat1}-\eqref{eq:wheat2} is \vspace{-1.5mm}
\begin{equation} \label{eq:msw}
 w(t)= \tline_t w_0 + \int_0^t \tline_{t-\tau} [\Ascr \nu f(\tau) - \nu \dot{f}(\tau)]\dd \tau, \vspace{-1.5mm}
\end{equation}
where $w_0=u_0-\nu f(0)\in L^2(0,1)$. Using \cite[Chapter 4, Corollary 2.5 and Theorem 3.5 (i)]{Pazy:1983} we get that $\Ascr w, \dot w \in C((0,T];L^2(0,1))$ and $w$ satisfies \eqref{eq:wheat1}-\eqref{eq:wheat2} for $t>0$. Changing the variable from $\tau$ to $s=t-\tau$ in the integral in \eqref{eq:msw}, then differentiating \eqref{eq:msw} and then changing the variable back from $s$ to $\tau = t-s$ we obtain that \vspace{-2mm}
$$ \dot w(t) = \Ascr \tline_t w_0 + \tline_t(\Ascr\nu f(0)-\nu\dot f(0)) + \int_0^t \tline_{t-\tau} [\Ascr \nu \dot f(\tau) - \nu \ddot{f}(\tau)]\dd \tau \vspace{-2mm}$$
for $t>0$. Since $\Ascr^2\tline_t w_0 \in C((0,T];L^2(0,1))$ (because $\tline$ is an analytic semigroup), again using \cite[Chapter 4, Theorem 3.5 (i)]{Pazy:1983} to deduce the regularity of the remaining terms on the right-side of the above expression, we get that $\Ascr \dot w \in C((0,T];L^2(0,1))$. So $\dot w \in C((0,T]; C^1[0,1])$. Using this and $w\in C((0,T];\Dscr(\Ascr))$ shown earlier, it follows that $\sigma w_x\in C((0,T];PC[0,1])$ and all  terms in \eqref{eq:wheat1} except $\sigma w_x$ and $\theta w_{xx}$ are in $C((0,T];PC^1[0,1])$. Hence $\theta w_{xx}\in C((0,T];PC[0,1])$, which implies that $\sigma w_x\in C((0,T]; PC^1[0,1])$. So all terms in \eqref{eq:wheat1} except $\theta w_{xx}$ are in $C((0,T];PC^1[0,1])$. Consequently $\theta w_{xx}\in C((0,T];PC^1[0,1])$ and so $w\in C((0,T]; PC^{(2),1}[0,1])$.

From the above discussion, in particular $u(x,t)= w(x,t) + \nu(x) f(t)$ and \eqref{eq:msw}, we get that $u$ defined in \eqref{eq:mildsoln} is in $C((0,T];PC^{(2),1}[0,1])\cap C^1((0,T];C^1[0,1])$ and satisfies \eqref{eq:heat1}-\eqref{eq:heat2} for $t>0$. The uniqueness of $w$, and so of $u$, follows easily from \cite[Chapter 4, Corollary 3.3]{Pazy:1983}.
\end{proof}


In the proof of the next lemma, we will denote the $j^{\rm th}$ component of a vector $v\in\rline^n$ by $[v]_j$. Recall $h=1/(n+1)$ and the operator $R_n$ from the notations in Section \ref{sec1}. Recall $\Iscr$, $\Jscr_n$ and $\Jscr_n^c$ defined below \eqref{eq:heat2} and \eqref{eq:Deltann}. \vspace{-1mm}

\begin{framed} \vspace{-2.5mm}
\begin{lemma}\label{lm:approxAq}
Fix $T>0$ and let $\rho\in[0,T)$. Consider $\xi \in C([\rho,T];PC^{(2),1}[0,1])$ satisfying $\alpha_0 \xi_x(0,t) + \beta_0 \xi(0,t)=0$. Define $f_\xi(t)=\alpha_1 \xi_x(1,t) + \beta_1 \xi(1,t)$. Let the points at which $\xi_{xx}(\cdot,t)$ is not differentiable be a subset of $\Iscr$ for all $t\in[\rho,T]$. Then for all $n\gg1$ and $p\in\{1,2\}$ there exists a $c$ independent of $\xi$ and $n$ such that  \vspace{-4mm}
\begin{align}
  &\sup_{t\in[\rho,T]}\|R_n \Ascr \xi(\cdot,t) - A_n R_n \xi(\cdot,t) - B_n f_\xi(t)\|_p \nonumber\\[-1ex]
  &\hspace{20mm}\leq\ c (h^{1-1/p}+1) \sup_{t\in[\rho,T]} \|\xi(\cdot,t)\|_{PC^{(2),1}[0,1]}. \label{eq:pwiseJ1one}\\[-6.9ex]\nonumber
\end{align}
\end{lemma}
\end{framed}
\vspace{-4mm}
\begin{proof}
For each $j\in \{1,2,\ldots n\}$, define $\Delta_n \xi(jh,t)$ by replacing $u(kh,t)$ with $\xi(kh,t)$ and $f(t)$ with $f_\xi(t)$ in \eqref{eq:Deltan1}-\eqref{eq:Deltann} and define $\nabla_n \xi(jh,t)$ by replacing $u(kh,t)$ with $\xi(kh,t)$ in \eqref{eq:nablan1}-\eqref{eq:nablanj}. Let $\delta_j=1$ if $j\in\Jscr_n^c$ and $\delta_j=0$ if $j\in\Jscr_n$.
Then for all $t\in[\rho,T]$, all $n\gg1$ and each $j\in \{1,2,\ldots n\}$, \vspace{-1mm}
\begin{align}
 |\xi_{xx}(jh,t) - \Delta_n \xi(jh,t)| &\,\leq c_1 h^{\delta_j}, \label{eq:xiDNest1}\\
 |\xi_x(jh,t)-\nabla_n \xi(jh,t)| &\,\leq c_2 h, \label{eq:xiDNest3} \\[-4.3ex]\nonumber
\end{align}
where $c_1=K \sup_{t\in[\rho,T]} \|\xi_{xx} (\cdot,t) \|_{PC^1}$, $c_2=K \sup_{t\in[\rho,T]} \|\xi_{xx}(\cdot,t) \|_{PC}$ and $K>0$ is independent of $n$ and $\xi$. These estimates follow from \eqref{eq:appr_uxxC} and \eqref{eq:appr_ux} and the discussion below them. For all $t\in[\rho,T]$, $n\gg1$ and $j\in\{1,2,\ldots n\}$, a simple calculation using \eqref{eq:xiDNest1}-\eqref{eq:xiDNest3} gives \vspace{-1mm}
\begin{equation} \label{eq:pwiseJ1}
  [R_n \Ascr \xi(\cdot,t) - A_n R_n \xi(\cdot,t) - B_n f_\xi(t)]_j \leq\ c_3 h^{\delta_j} \sup_{t\in[\rho,T]} \|\xi(\cdot,t)\|_{PC^{(2),1}[0,1]} \vspace{-1mm}
\end{equation}
for some $c_3>0$ independent of $n$ and $\xi$. Recalling that the cardinality of $\Jscr_n$ is bounded by a constant independent of $n$, \eqref{eq:pwiseJ1one} follows easily from \eqref{eq:pwiseJ1}. \vspace{-2mm}
\end{proof}

\subsection{Convergence results} \label{sec3.2} \vspace{-1mm}

\ \ \ In this section we present our main convergence results, Theorem \ref{th:semidisc_conv} and Corollary \ref{cor:semidisc_conv}. Theorem \ref{th:semidisc_conv} is used to establish Theorem \ref{thm:finalmainresult} (the main result of this paper) and  Corollary \ref{cor:semidisc_conv} is used to establish Proposition \ref{prop:anjconvj}. Recall $A_n$ from \eqref{eq:semi-disc} and  $h=1/(n+1)$.  We will need the next proposition, which gives a growth bound independent of $n$ for the semigroup $e^{A_n t}$, to establish the convergence results. Recall the notation $\|v\|_p$ and $\|v\|_\infty$ for $v\in\rline^n$. \vspace{-1mm}

\begin{framed} \vspace{-2mm}
\begin{proposition}\label{pr:eigenA}
Consider the $n^{\rm th}$-order ODE \vspace{-1mm}
\begin{equation} \label{eq:lmODE}
 \dot v(t) =  A_n v(t) + \eta(t), \qquad v(0)=v_0, \vspace{-1mm}
\end{equation}
with a forcing term $\eta\in L^\infty([0,\infty);\rline^n)$. Then there exist constants $M, \omega>0$ independent of $n$ such that the solution $v$ to the ODE satisfies \vspace{-1mm}
\begin{equation} \label{eq:lmbnd}
 \|v(t)\|_2 \leq M e^{\omega t} \|v_0\|_2+\sqrt{h} \int_0^t M e^{\omega (t-\tau)} \| \eta(\tau)\|_1\dd\tau \FORALL t\geq0. \vspace{-4mm}
\end{equation}
\end{proposition}
\end{framed}
\vspace{-5mm}

\begin{proof}
For $z= [z_1 \ \ z_2 \ \cdots \ z_n]^\top\in\rline^n$, define \vspace{-3mm}
\begin{equation} \label{eq:defnD}
 \|\DD z\|^2_2 = \sum_{j=1}^{n-1} \frac{(z_{j+1}-z_j)^2}{h^2}. \vspace{-3mm}
\end{equation}
For each $p,k\in\{1,2,\ldots n\}$, we have \vspace{-1.5mm}
$$ z^2_k = z_p^2 + (-1)^{\mu}\sum_j (z_{j+1}^2-z_j^2) \leq z_p^2 + \sum_j [h \epsilon^{-1}(z_{j+1}+z_j)^2 + h^{-1}\epsilon (z_{j+1}-z_j)^2], \vspace{-2.5mm}$$
where the summation is over the integers $j$ between $p$ and $k$, $\mu=1$ if $p>k$ and $\mu=0$ if $p\leq k$ and $\epsilon>0$. To get the inequality in this expression we have used the Young's inequality $ab \leq h^{-1}\epsilon a^2 + h\epsilon^{-1} b^2$. It follows from the above expression that \vspace{-1.5mm}
$$ z^2_k \leq z_p^2 + 4h \epsilon^{-1} \|z\|_2^2 + h\epsilon \|\DD z\|_2^2.  \vspace{-1.5mm}$$
Summing both sides of the above inequality over $p\in\{1,2,\ldots n\}$ and then  taking supremum over $k\in\{1,2,\ldots n\}$ we get \vspace{-1.5mm}
\begin{equation} \label{eq:sobemb}
 \|z\|^2_\infty \leq h\epsilon \|\DD z\|^2_2 + (2h+ 4h \epsilon^{-1}) \|z\|^2_2. \vspace{-2mm}
\end{equation}

Recall $L_n,D_n, \Theta_n, \Sigma_n,\Lambda_n \in\rline^{n\times n}$ and $r_0,r_1,q_0\in\rline$ introduced below \eqref{eq:semi-disc}. Note that $\|D_n z\|_2^2 = \|\DD z\|_2^2+q_0^2z_1^2$. Let $\Gamma_n = {\rm diag}\bbm{(1-r_0) & 1 &1 \ \cdots \ 1 & (1-r_1)} \in \rline^{n\times n}$. A simple calculation gives \vspace{-1.5mm}
\begin{equation} \label{eq:vLnv_expr}
    z^\top \Gamma_n^{-1}L_n z  = \left[q_0^2 z_1^2 + \frac{(3r_0-1)z_1^2}{h^2(1-r_0)} +\frac{(3r_1-1)z_n^2}{h^2(1-r_1)} \right]- z^\top D_n^\top D_n z. \vspace{-1.5mm}
\end{equation}
Note that $h^2 q_0^2 + (3r_0-1)/(1-r_0) = h\alpha_0\beta_0/ (\alpha_0-h\beta_0)^2$ and $(3r_1-1)/(1-r_1)= -h\beta_1/ (\alpha_1+h\beta_1)$. So for all $n\gg1$ there exists a constant $C_r>0$ independent of $h$ such that the coefficients of $z_1^2$ and $z_n^2$ in the above equation are less than $C_r/h$. Using this, the inequality $\|D_n z\|^2_2\geq \|\DD z\|^2_2$ and \eqref{eq:sobemb} to bound the terms on the right-side of \eqref{eq:vLnv_expr} we get \vspace{-1.5mm}
\begin{equation}\label{eq:vLv_est}
  z^\top \Gamma_n^{-1}L_n z  \leq 2C_r (2+ 4 \epsilon^{-1}) \|z\|^2_2 - (1-2C_r\epsilon)\|D_n z\|_2^2. \vspace{-1.5mm}
\end{equation}
For all $n\gg1$, note that we have $(1-r_0),(1-r_1)\in[0.5,1]$ and so the absolute value of each entry of the diagonal matrix $\Gamma_n^{-1} \Theta_n^{-1}\Sigma_n$ can be bounded by a constant $C_q$ independent of $n$. Now using the Young's inequality we get \vspace{-1.5mm}
\begin{equation}\label{eq:vSv_est}
 z^\top \Gamma_n^{-1}\Theta_n^{-1}\Sigma_n D_n z \leq C_q \epsilon^{-1} \|z\|_2^2 + C_q \epsilon \|D_n z\|_2^2. \vspace{-1.5mm}
\end{equation}
The absolute value of each entry of the diagonal matrix $\Gamma_n^{-1} \Theta_n^{-1} \Lambda_n$ can be bounded by $C_l>0$ independent of $n$. It follows that \vspace{-1.5mm}
\begin{equation}\label{eq:vLPv_est}
 |z^\top \Gamma_n^{-1} \Theta_n^{-1} \Lambda_n z| \leq C_l \|z\|_2^2. \vspace{-1.5mm}
\end{equation}
Note that $A_n=\Theta_n L_n + \Sigma_n D_n + \Lambda_n$. It now follows from the estimates in \eqref{eq:vLv_est}, \eqref{eq:vSv_est} and \eqref{eq:vLPv_est} with $\epsilon>0$ sufficiently small that for all $n\gg1$ there exists $c_1>0$ and $c_2>0$ independent of $n$ such that \vspace{-1.5mm}
\begin{equation} \label{eq:Anest}
 z^\top  \Gamma_n^{-1}\Theta_n^{-1} A_n z \leq c_1 \|z\|_2^2 - c_2 \|D_n z\|_2^2. \vspace{-2mm}
\end{equation}

Along the solution of \eqref{eq:lmODE}, define $W(t)=v(t)^\top \Gamma_n^{-1}\Theta_n^{-1} v(t)/2$ for all $t\geq0$. Clearly, $\dot W(t) = v(t)^\top \Gamma_n^{-1}\Theta_n^{-1} A_n v(t) + v(t)^\top \Gamma_n^{-1} \Theta_n^{-1} \eta(t)$. It follows from \eqref{eq:Anest} that \vspace{-1mm}
\begin{equation}\label{eq:vtAnv}
 \dot W(t) \leq c_1 \|v(t)\|_2^2 - c_2 \|D_n v(t)\|_2^2 + v(t)^\top \Gamma_n^{-1}\Theta_n^{-1} \eta(t).
\end{equation}
Applying H\"{o}lder's inequality and Young's inequality, and using the fact that the entries of $\Gamma_n^{-1}\Theta_n^{-1}$ are bounded by a constant independent of $n$ (for $n$ large), we get that there exists $c_3>0$ such that \vspace{-2mm}
$$ v^\top(t)\Gamma_n^{-1}\Theta_n^{-1}\eta(t) \leq  \frac{\|v(t)\|_\infty^2}{h} + c_3 h\|\eta(t)\|_1^2. \vspace{-2mm} $$
Using \eqref{eq:sobemb} with $\epsilon = c_2$ to bound $\|v(t)\|_\infty^2/h$ in the above expression we get \vspace{-1.5mm}
$$ v^\top(t)\Gamma_n^{-1}\Theta_n^{-1}\eta(t) \leq (2+4c_2^{-1}) \|v(t)\|_2^2 + c_2\|\DD v(t)\|_2^2 +  c_3 h\|\eta(t)\|_1^2. \vspace{-1.5mm}$$
Using the above estimate in \eqref{eq:vtAnv}, recalling that $\|\DD v(t)\|_2^2\leq \|D_n v(t)\|_2^2$ and noting that for all $n\gg1$ there exist $k_1, k_2>0$ independent of $n$ such that $k_1^2 \|v(t)\|_2^2 \leq W(t) \leq k_2^2 \|v(t)\|_2^2$, we get $\dot W(t) \leq  k W(t) + c_3 h \|\eta(t)\|_1^2$ for some $k>0$.
Therefore \vspace{-2.5mm}
$$ W(t)\leq e^{k t} W(0) + c_3 h\int_0^t e^{k (t-\tau)} \|\eta(\tau)\|_1^2 \dd\tau. \vspace{-2.5mm} $$
Consequently $\|v(t)\|^2_2 \leq (k_2/k_1)^2 e^{k t}\|v(0)\|^2_2+
(c_3 h/k_1^2) \int_0^t e^{k (t-\tau)} \|\eta(\tau)\|_1^2\dd\tau$ for all $t\geq0$ and $n\gg1$, which implies the estimate in \eqref{eq:lmbnd}. \vspace{-1mm}
\end{proof}

The next theorem shows that the solution of the semi-discrete approximation \eqref{eq:semi-disc} converges to the solution of the PDE \eqref{eq:heat1}-\eqref{eq:heat2} for some initial states and inputs. Recall the notation $\|v\|_{2d}$ for $v\in \rline^n$ and the operators $R_n$ and $S_n$ from the notations in Section \ref{sec1}. In the proof we will often use the fact that $\|S_n z\|_{L^2(0,1)}=\|z\|_{2d}$ for each $z\in\rline^n$. \vspace{-2.5mm}

\begin{framed} \vspace{-2.5mm}
\begin{theorem}\label{th:semidisc_conv}
Fix $T>0$. Consider $u_0\in C[0,1]$, a function $f\in C^3[0,T]$ and a sequence of functions $\{f_n\}_{n=1}^\infty$ in $C^3[0,T]$ such that \vspace{-2mm}
\begin{equation} \label{eq:fn_limit}
 \lim_{n\to\infty}\|f_n-f\|_{C^1[0,T]}=0. \vspace{-2mm}
\end{equation}
Let $u$ be the solution of \eqref{eq:heat1}-\eqref{eq:heat2} on the time interval $[0,T]$ with initial state $u_0$ and input $f$. Let $v_n$ be the solution of the $n^{\rm th}$-order semi-discrete system \eqref{eq:semi-disc} on the time interval $[0,T]$ with initial state $R_n u_0$ and input $f_n$. Then \vspace{-2mm}
\begin{equation}\label{eq:semdisconv2}
  \lim_{n\to\infty}\sup_{t\in[0,T]}\|u(\cdot,t) - S_n v_n(t)\|_{L^2(0,1)}= 0. \vspace{-6mm}
\end{equation}
\end{theorem}
\end{framed}
\vspace{-6.5mm}
\begin{proof}
From Lemma \ref{lm:reg} we get $u\in C((0,T];PC^{(2),1}[0,1])$ and $\dot u\in C((0,T];C^1[0,1])$. Using this in \eqref{eq:heat1}
it is easy to see that the points at which $u_{xx}(\cdot,t)$ is not differentiable is a subset of $\Iscr$ for all $t\in(0,T]$. First we will show that \vspace{-2mm}
\begin{equation} \label{eq:initcont}
 \lim_{t\to0} \sup_{n\in\nline} \|v_n(t)-R_n u_0\|_{2d}=0. \vspace{-2mm}
\end{equation}
Let $e_n(t) = R_n u(\cdot,t) - v_n(t) + R_n \nu (f_n(t)-f(t))$. Then \vspace{-1.5mm}
\begin{equation} \label{eq:useexp}
 u(\cdot,t)-S_n v_n(t) = [u(\cdot,t)-S_n R_n u(\cdot,t)] + S_n e_n(t) - [S_nR_n\nu(f_n(t)-f(t))]. \vspace{-1.5mm}
\end{equation}
For any $\rho\in(0,T)$ we will show that \vspace{-2mm}
\begin{equation} \label{eq:enl2}
 \lim_{n\to\infty}\sup_{t\in[\rho,T]}\|e_n(t)-e^{A_n (t-\rho)} e_n(\rho) \|_{2d} =0. \vspace{-2mm}
\end{equation}
The equations \eqref{eq:initcont} and \eqref{eq:enl2} imply \eqref{eq:semdisconv2}. Indeed, fix $\epsilon>0$. Using $u\in C([0,T]; L^2(0,1))$ and \eqref{eq:initcont} choose $\rho_\epsilon$ such that $\|u(\cdot,t)-u_0\|_{L^2(0,1)}\leq \epsilon$ and $\|S_n v_n(t)-S_nR_n u_0\|_{L^2(0,1)}<\epsilon$ for all $t\in[0,\rho_\epsilon]$ and $n\in\nline$. Since $u_0\in C[0,1]$, there exists $\tilde n_\epsilon$ such that $\|u_0 - S_n R_n u_0\|_{L^2(0,1)}<\epsilon$ for $n\geq\tilde n_\epsilon$. From these inequalities we get \vspace{-1.5mm}
\begin{equation} \label{eq:pfrho}
 \sup_{t\in[0, \rho_\epsilon]} \|u(\cdot,t)-S_n v_n(t)\|_{L^2(0,1)} <3\epsilon \vspace{-2mm}
\end{equation}
for all $n\geq\tilde n_\epsilon$. Next choose $n_\epsilon>\tilde n_\epsilon$ such that for each $n\geq n_\epsilon$, \vspace{-1.5mm}
\begin{align}
 &\sup_{t\in[0,T]}\|S_nR_n\nu(f_n(t)-f(t))\|_{L^2} <\epsilon, \quad \sup_{t\in[\rho_\epsilon, T]} \|u(\cdot,t) - S_n R_n u(\cdot,t)\|_{L^2}<\epsilon, \label{eq:snrn}\\
 &\hspace{40mm}\sup_{t\in[\rho_\epsilon,T]} \|e_n(t)-e^{A_n (t-\rho_\epsilon)}e_n(\rho_\epsilon)\|_{2d} <\epsilon. \label{eq:ent-} \\[-4.9ex]\nonumber
\end{align}
The existence of such an $n_\epsilon$ is guaranteed by  \eqref{eq:fn_limit}, the uniform continuity of $u$ on $[0,1]\times[\rho_\epsilon,T]$ (note that $u\in C([\rho_\epsilon,T];C[0,1])$) and \eqref{eq:enl2}. Then for $n\geq n_\epsilon$, using \eqref{eq:pfrho} and \eqref{eq:snrn} in \eqref{eq:useexp} with $t=\rho_\epsilon$ we get $\|e_n(\rho_\epsilon)\|_{2d} <5\epsilon$. From this, via Proposition \ref{pr:eigenA}, we get $\|e^{A_n (t-\rho_\epsilon)} e_n(\rho_\epsilon) \|_{2d}\leq 5\epsilon M e^{\omega T}$ for $t\in[\rho_\epsilon,T]$. This estimate and \eqref{eq:ent-} give $\|S_n e_n(t)\|_{L^2(0,1)} <\epsilon+5\epsilon M e^{\omega T}$ for $t\in[\rho_\epsilon,T]$. Using this and \eqref{eq:snrn} in \eqref{eq:useexp} we get \vspace{-4mm}
$$ \sup_{t\in[\rho_\epsilon,T]} \|u(\cdot,t)-S_n v_n(t)\|_{L^2(0,1)} <
  3\epsilon + 5\epsilon Me^{\omega T}\vspace{-3mm} $$
for all $n\geq n_\epsilon$. Since \eqref{eq:pfrho} and the above estimate can be established for any $\epsilon>0$, it follows that \eqref{eq:semdisconv2} holds. We will now complete the proof of the theorem by showing that \eqref{eq:initcont} and \eqref{eq:enl2} hold. \vspace{-1mm}

We will first prove \eqref{eq:initcont}. Recall $\nu$ defined below \eqref{eq:mildsoln}. Fix $\epsilon>0$ and $\tilde u_0\in C^3[0,1]$ such that $\tilde u_0$ has a compact support in $(0,1)$ and $\|u_0 - \nu f(0) - \tilde u_0\|_{L^2(0,1)}< \epsilon$. Define $z_n(t)=v_n(t)-R_n\nu f_n(t)-R_n \tilde u_0$. Then \vspace{-1.5mm}
\begin{equation} \label{eq:znode}
 \dot z_n(t) = A_n z_n(t) +  \eta_n(t), \qquad z_n(0)=R_n (u_0 - \nu f_n(0) - \tilde u_0), \vspace{-1.5mm}
\end{equation}
where $\eta_n(t)= (B_n+A_n R_n \nu) f_n(t) + A_n R_n \tilde u_0 - R_n\nu \dot f_n(t)$. Since $u_0 - \nu f(0) - \tilde u_0 \in C[0,1]$ and \eqref{eq:fn_limit} holds, there exists $\tilde t_\epsilon>0$ and $n_\epsilon$ such that for $t\leq \tilde t_\epsilon$ and $n\geq n_\epsilon$, \vspace{-1.5mm}
\begin{equation} \label{eq:trbea}
\|R_n\nu(f_n(t)-f_n(0))\|_{2d}\leq \epsilon, \quad \|z_n(0)\|_{2d}<2\epsilon. \vspace{-1.5mm}
\end{equation}
Taking $p=1$ and applying Lemma \ref{lm:approxAq} with $\xi(\cdot,t)=\tilde u_0$ and $\xi(\cdot,t)=\nu$ we get, for some $c_1>0$ and all $n\gg1$, that $\|A_n R_n \tilde u_0\|_1 \leq \|R_n\Ascr\tilde u_0\|_1 + c_1 $ and \vspace{-1.5mm}
\begin{equation}  \label{eq:nulemma}
 \|A_n R_n \nu + B_n\|_1 \leq \|R_n\Ascr\nu\|_1 + c_1. \vspace{-1.5mm}
\end{equation}
Using this, the fact that $\|R_n\nu\|_1$, $\|R_n\Ascr\nu\|_1$ and $\|R_n\Ascr\tilde u_0\|_1$ can each be bounded by $c_2 h^{-1}$ for some $c_2>0$ and \eqref{eq:fn_limit}, it follows that $\|\eta_n(t)\|_1< c_3 h^{-1}$ for some $c_3>0$, each $t\in[0,T]$ and all $n$. So applying Proposition \ref{pr:eigenA} to \eqref{eq:znode} we can get \vspace{-1.5mm}
$$ \|z_n(t)\|_{2d} \leq M e^{\omega t} \|z_n(0)\|_{2d} + c t \FORALL t\in[0,T] \vspace{-1.5mm}$$
and some $c>0$ that depends on $\tilde u_0$ and so on $\epsilon$. The above inequality and \eqref{eq:trbea} guarantee the existence of $t_\epsilon<\tilde t_\epsilon$ such that $\|z_n(t)\|_{2d}<4M\epsilon$ for $n\geq n_\epsilon$ and $t\in[0,t_\epsilon]$. From this, the expression $v_n(t)-R_n u_0 = z_n(t) - z_n(0) + R_n\nu(f_n(t)-f_n(0))$ and \eqref{eq:trbea}, it now follows easily that
$\|v_n(t)-R_n u_0\|_{2d}<7M\epsilon$ for $n\geq n_\epsilon$ and $t\in[0,t_\epsilon]$. Finally, using $\lim_{t\to0} \|v_n(t)-R_n u_0\|_{2d}=0$ for each $n$ and redefining $t_\epsilon$ if needed, we get $\|v_n(t)-R_n u_0\|_{2d}<7M\epsilon$ for all $n\in\nline$ and $t\in[0,t_\epsilon]$. Since this estimate can be established for any $\epsilon>0$, we conclude that \eqref{eq:initcont} holds. \vspace{-1mm}

We will now complete this proof by establishing \eqref{eq:enl2}. Recall $e_n$ defined below \eqref{eq:initcont}. From \eqref{eq:heat1} and \eqref{eq:semi-disc} we get \vspace{-1.5mm}
\begin{equation} \label{eq:einit}
  \dot{e}_n(t) = A_n e_n(t) + \eta^1_n(t)+\eta^2_n(t)+\eta^3_n(t), \vspace{-1.5mm}
\end{equation}
where $\eta_n^1(t)= [A_n R_n \nu + B_n](f(t)-f_n(t))$, $\eta_n^2(t)=R_n \Ascr u(\cdot,t) - A_n R_n u(\cdot,t)-B_n f(t)$ and $\eta_n^3(t)= R_n \nu (\dot{f}_n(t) - \dot{f}(t))$.
From \eqref{eq:nulemma}, the estimate for $\|R_n\Ascr\nu\|_1$ given below it and \eqref{eq:fn_limit} we get that $\lim_{n\to\infty} \sup_{t\in[0,T]} h \|\eta_n^1(t)\|_1 =0$. Since $u\in C([\rho,T];PC^{(2),1}[0,1])$ for any given $\rho\in(0,T)$, see Lemma \ref{lm:reg}, applying Lemma \ref{lm:approxAq} with $p=1$ and $\xi=u$ we get $\sup_{t\in[\rho,T]}\|\eta_n^2(t)\|_1 \leq \tilde c(\rho)$ for all $n\gg1$ and a constant $\tilde c(\rho)$ independent of $n$. So $\lim_{n\to\infty}\sup_{t\in[\rho,T]} h \|\eta_n^2(t)\|_1=0$. Finally, it follows from \eqref{eq:fn_limit} and the estimate for $\|R_n\nu\|_1$ below \eqref{eq:nulemma} that $\lim_{n\to\infty}\sup_{t\in[0,T]} h \|\eta_n^3(t)\|_1\\=0$. In summary we get that for any $\rho>0$, \vspace{-2mm}
\begin{equation} \label{eq:eta_conv}
 \lim_{n\to\infty} h \sup_{t\in [\rho,T]}\|\eta_n^1(t)+ \eta_n^2(t)+\eta_n^3(t)\|_1 =0. \vspace{-2.5mm}
\end{equation}
Applying Proposition \ref{pr:eigenA} to \eqref{eq:einit} and using the above estimate, \eqref{eq:enl2} follows.
\end{proof}

The following corollary to Theorem \ref{th:semidisc_conv} gives a pointwise convergence result. Recall the notation $v_n(t)=[v_{n,1}(t) \ \ v_{n,2}(t) \ \cdots \ v_{n,n}(t)]^\top$ and $q_0$ defined above \vspace{-2mm} \eqref{eq:vn2flat}.

\begin{framed} \vspace{-2mm}
\begin{corollary}\label{cor:semidisc_conv}
Suppose that in Theorem \ref{th:semidisc_conv} we let $u_0\in\Dscr(\Ascr^\infty)$, $f, f_n\in C^\infty[0,T]$ with $\lim_{n\to\infty}\|f_n-f\|_{C^k[0,T]}=0$ and  $f_n^{(k)}(0)=0$ for all integers $k\geq0$ and $n\geq1$. Then \vspace{-1.5mm}
\begin{equation} \label{eq:semdisconv8}
  \lim_{n\to\infty}\sup_{t\in[0,T]}|u(0,t) - v_{n,1}(t)| + |u_x(0,t) - q_0 v_{n,1}(t)| = 0. \vspace{-4mm}
\end{equation}
\end{corollary}
\end{framed}
\vspace{-6mm}
\begin{proof}
Note that $f^{(k)}(0)=0$ for all integers $k\geq0$. Using \cite[Chapter 4, Theorem 3.5 (ii)]{Pazy:1983} instead of \cite[Chapter 4, Theorem 3.5 (i)]{Pazy:1983} in the proof of Lemma \ref{lm:reg} it follows that the solution $u$ of \eqref{eq:heat1}-\eqref{eq:heat2} with initial state $u_0$ and input $f$ is in $C([0,T]; PC^{(2),1}[0,1])$. Furthermore, $\dot u \in C([0,T];C^1[0,1])$ and $u$ satisfies \eqref{eq:heat1}-\eqref{eq:heat2} for $t\geq0$. (The result \cite[Chapter 4, Theorem 3.5 (ii)]{Pazy:1983} enables us to replace $(0,T]$ in Lemma \ref{lm:reg} with $[0,T]$). Let $\bar u$ denote the solution of \eqref{eq:heat1}-\eqref{eq:heat2} with initial state $\Ascr u_0$ and input $\dot f$. Then, like $u$ above, we get $\bar u \in C([0,T]; PC^{(2),1}[0,1])\cap C^1([0,T]; C^1[0,1])$ satisfies \eqref{eq:heat1}, $\alpha_0 \bar u_x(1,t) + \beta_0 \bar u(1,t)=0$ and $\alpha_1 \bar u_x(1,t) + \beta_1 \bar u(1,t)=\dot f(t)$ for $t\in[0,T]$. Differentiating \eqref{eq:mildsoln} it is easy to check that the resulting formula for $\dot u$ is in fact the convolution formula for $\bar u$, i.e. $\bar u =\dot u$ and $\dot u$ has the same regularity as $\bar u$. \vspace{-2mm}

Recall $e_n(t)$ defined below \eqref{eq:initcont} and $\eta_n^1(t)$, $\eta_n^2(t)$, $\eta_n^3(t)$ introduced in \eqref{eq:einit}. Let $\eta_n(t) \!=\!\eta_n^1(t)+\eta_n^2(t)+\eta_n^3(t)$. Clearly $e_n(0)=0$ and $\dot e_n(0)=R_n \Ascr u_0 - A_n R_n u_0$. Since $u\in C([0,T];PC^{(2),1}[0,1])$, we can take $\rho=0$ in the proof of Theorem \ref{th:semidisc_conv} to conclude from \eqref{eq:enl2} that \vspace{-4mm}
\begin{equation} \label{eq:en_rho0}
  \lim_{n\to\infty}\sup_{t\in [0,T]} \|e_n(t)\|_{2d} =0. \vspace{-3.5mm}
\end{equation}
Using $\dot u\in C([0,T];PC^{(2),1}[0,1])$ and $\alpha_1 \dot u_x(1,t) + \beta_1 \dot u(1,t)=\dot f(t)$, we can replicate the arguments used to derive \eqref{eq:eta_conv} to conclude that $\lim_{n\to\infty} h \sup_{t\in [0,T]}\|\dot\eta_n(t)\|_1 =0$. Consequently, taking the derivative of \eqref{eq:einit} and applying Proposition \ref{pr:eigenA} to the resulting equation, and inferring using Lemma \ref{lm:approxAq} that $\lim_{n\to\infty}\|\dot e_n(0)\|_{2d}=0$, we get \vspace{-1mm}
\begin{equation} \label{eq:doten_rho0}
 \lim_{n\to\infty}\sup_{t\in [0,T]} \|\dot e_n(t)\|_{2d} =0. \vspace{-1mm}
\end{equation}
Using Lemma \ref{lm:approxAq} with $p=2$, it is easy to mimic the steps used to derive \eqref{eq:eta_conv} to show that $\lim_{n\to\infty} \sup_{t\in[0,T]} \|\eta_n(t)\|_{2d}=0$. From this, \eqref{eq:doten_rho0}, \eqref{eq:einit} and $e_n, \dot e_n, \eta_n\in C([0,T];\rline^n)$ we get \vspace{-1mm}
\begin{equation}\label{eq:estAn3}
  \lim_{n\to\infty}\sup_{t\in [0,T]} \|A_n e_n(t)\|_{2d} =0. \vspace{-1mm}
\end{equation}
The entries of the diagonal matrix $\Gamma_n^{-1}\Theta_n^{-1}$ are bounded by a constant independent of $n$. So it follows from \eqref{eq:Anest} that there exists $c_1>0$ such that \vspace{-1mm}
\begin{equation} \label{eq:Dnen2d}
  \|D_n e_n(t)\|_{2d} \leq c_1(\|e_n(t)\|_{2d}+\|A_n e_n(t)\|_{2d}) \vspace{-1mm}
\end{equation}
for all $n\gg1$ and $t\in[0,T]$. From the definition of $A_n$ (see below \eqref{eq:semi-disc}) we get $L_n e_n = \Theta_n^{-1}(A_n - \Sigma_n D_n - \Lambda_n)e_n$. There exists a constant $c>0$ such that the absolute value of each entry  of the  diagonal matrices $\Theta_n^{-1}$, $\Sigma_n$ and $\Lambda_n$ is less than $c$. From this and \eqref{eq:Dnen2d} we get that there exists $c_2>0$ such that  \vspace{-1mm}
\begin{equation} \label{eq:Lnen2d}
 \|L_n e_n(t)\|_{2d}\leq c_2(\|e_n(t)\|_{2d}+\|A_n e_n(t)\|_{2d}) \vspace{-1mm}
\end{equation}
for all $n\gg1$ and $t\in[0,T]$. Let $e_{n,k}$ and $[D_n e_n]_k$ denote the $k^{\rm th}$-entry of $e_n$ and $D_n e_n$, respectively. It is easy to check using the definitions of $r_0$, $q_0$ and $D_n$ that $(-2+4 r_0) e_{n,1}+(1-r_0)e_{n,2} = h(1-r_0)([D_n e_n]_2 - [D_n e_n]_1)$ and $r_0\in[0,0.5]$ for $n\gg1$. Now a simple calculation shows that $\|L_n e_n\|^2_2 \geq (1-r_0)^2 \|\DD D_n e_n\|^2_2$, where $\DD$ is as defined in \eqref{eq:defnD}. From this and \eqref{eq:Lnen2d} we get that  \vspace{-1mm}
\begin{equation} \label{eq:DDnen2d}
  \sqrt{h} \|\DD D_n e_n(t)\|_2 \leq c_3(\|e_n(t)\|_{2d}+\|A_n e_n(t)\|_{2d})
\end{equation}
for some $c_3>0$, for all $n\gg1$ and $t\in[0,T]$. Using \eqref{eq:Dnen2d} and \eqref{eq:DDnen2d} and the fact $\|D_n e_n(t)\|^2_2\geq \|\DD e_n(t)\|^2_2$, by applying \eqref{eq:sobemb} we get that there exists a constant $c_4>0$ such that \vspace{-2mm}
\begin{equation}\label{eq:estDen8}
\| e_n(t) \|_\infty + \|D_n e_n(t) \|_\infty  \leq c_4(\|e_n(t)\|_{2d}+\|A_n e_n(t)\|_{2d}) \vspace{-1mm}
\end{equation}
for all $n\gg1$ and $t\in[0,T]$. Observe that from the definition of $e_n$ we have \vspace{-1mm}
\begin{align}
 u(0,t) - v_{n,1}(t) &\,= u(0,t) - u(h,t) + e_{n,1}(t) + \nu(h) (f(t)-f_n(t)), \label{eq:inff1}\\
 u_x(0,t) - q_0v_{n,1}(t) &\, = u_x(0,t)-q_0u(h,t) + q_0 e_{n,1}(t) + q_0 \nu(h)(f(t)-f_n(t)). \label{eq:inff2}
\end{align}
Since $u\in C([0,T];PC^{(2),1}[0,1])$ we get $\lim_{h\to0}|u(0,t)-u(h,t)|=0$ and, using also \eqref{eq:nablan1} and \eqref{eq:appr_ux}, we get  $\lim_{h\to0}|u_x(0,t)-q_0 u(h,t)|=0$, uniformly in $t\in[0,T]$. From the definition of $\nu$ it follows that
$\lim_{h\to0} \nu(h)= \lim_{h\to0} q_0\nu(h)=0$. Finally, from \eqref{eq:en_rho0}, \eqref{eq:estAn3} and \eqref{eq:estDen8} we get $\lim_{n\to\infty} e_{n,1}(t)= \lim_{n\to\infty} q_0e_{n,1}(t)=0$ uniformly in $t\in[0,T]$. These limits, together with \eqref{eq:inff1}-\eqref{eq:inff2}, imply \eqref{eq:semdisconv8}. \vspace{-5mm}
\end{proof}


\section{Initial and final states for semi-discrete systems} \label{sec4}
\setcounter{equation}{0} 

\ \ \ In the approach described in Section \ref{sec2.1} for solving Problem \ref{prob:motionplan}, i.e. the problem of transferring \eqref{eq:heat1}-\eqref{eq:heat2} from an initial state $u_0$ to a final state $u_T$ over the time interval $[0,T]$, we first choose a $y\in G_\alpha[0,T]$ with $1<\alpha<2$ which satisfies \eqref{eq:y_endpts1}-\eqref{eq:y_endpts2}. We then define $y_n=y$ to obtain $v_{n,1}$ from \eqref{eq:yn} and using it in \eqref{eq:vn2flat}-\eqref{eq:fnflat} we get $f_n$. The limit of $f_n$ as $n$ goes to infinity is the input $f$ which solves Problem \ref{prob:motionplan}. In this approach, $y_n=y$ determines the initial state $v_n(0)=[v_{n,1}(0) \ \ v_{n,2}(0) \ \cdots \ v_{n,n}(0)]^\top$ and the final state $v_n(T)=[v_{n,1}(T) \ \ v_{n,2}(T) \ \cdots \ v_{n,n}(T)]^\top$ for the $n^{\rm th}$-order system \eqref{eq:semi-disc} via \eqref{eq:yn} and \eqref{eq:vn2flat}-\eqref{eq:vnjflat}. In this section, we will prove that \vspace{-1mm}
\begin{equation}\label{eq:vn0Tconv}
  \lim_{n\to\infty} \|R_n u_0 - v_n(0)\|_{2d}=0, \qquad \lim_{n\to\infty} \|R_n u_T - v_n(T)\|_{2d}=0.  \vspace{-1mm}
\end{equation}
These limits are used to prove our main result Theorem \ref{thm:finalmainresult}. We first present a lemma and a proposition needed to prove \eqref{eq:vn0Tconv}. Recall the sets $\Iscr$, $\Jscr_n$, $\Jscr_n^c$ defined below \eqref{eq:heat2} and \eqref{eq:Deltann} and the notation $[v]_j$ from above Lemma \ref{lm:approxAq}. \vspace{-1.5mm}

\begin{framed} \vspace{-3mm}
\begin{lemma}\label{lm:lmestAku}
Let $u_0\in PC^{(2),1}[0,1]$ be as in the statement of Problem \ref{prob:motionplan}. Then there exists a constant $\Gamma$ independent of $n$ and $k$ such that for all $n\gg1$, $k\geq 0$ and $j\in\{1,2,\ldots n-1\}$ we have \vspace{-1.5mm}
\begin{equation}\label{eq:AkAnknotJ}
  |[R_n \Ascr^{k+1} u_0 - A_n R_n \Ascr^k u_0]_j| \leq h^{\delta_j} \Gamma^{k+2}(k+2)!. \vspace{-1.5mm}
\end{equation}
Here $\delta_j=1$ if $j\in \Jscr_n^c$ and $\delta_j=0$ if $j\in \Jscr_n$. \vspace{-3mm}
\end{lemma}
\end{framed}
\vspace{-6mm}
\begin{proof}
From the definition of $\Ascr$, for each $k\geq0$ we have \vspace{-1mm}
\begin{equation} \label{eq:Akk+1}
 [\Ascr ^k u_0]_{xx} = \theta^{-1}\big(\Ascr^{k+1} u_0 - \sigma [\Ascr^k u_0]_x - \lambda \Ascr^k u_0\big). \vspace{-1mm}
\end{equation}
Let $\beta\in\cline$ be in the resolvent set of $\Ascr$. Then $(\Ascr-\beta I)^{-1}$ is a bounded linear operator from $L^2(0,1)$ to $H^2(0,1)$. Using the identity $\Ascr^k u_0 = (\Ascr-\beta I)^{-1}[\Ascr^{k+1} u_0 - \beta \Ascr^k u_0]$ which holds for $k\geq1$ and \eqref{eq:probest} we get that there exists a $c_1>0$ such that
$\|\Ascr^k u_0\|_{H^2(0,1)} \leq c_1^{k+1}(k+1)!$ for $k\geq1$.
From this estimate, applying Morrey's inequality, we get that
$\Ascr^k u_0\in C^1[0,1]$ and there exists a $c>0$ such that \vspace{-1mm}
\begin{equation} \label{eq:Aku01}
 \|\Ascr^k u_0\|_{C^1[0,1]} \leq c\,c_1^{k+1}(k+1)!. \vspace{-1mm}
\end{equation}
Using this in \eqref{eq:Akk+1} we can conclude that $[\Ascr ^k u_0]_{xx}\in PC[0,1]$ and $\|[\Ascr ^k u_0]_{xx}\|_{PC} \leq c_2^{k+2}(k+2)!$ for some $c_2>0$ and all $k\geq1$. This and \eqref{eq:Aku01} imply that $[\Ascr ^k u_0]_x\in PC^1[0,1]$ and \vspace{-1mm}
\begin{equation} \label{eq:Aku02}
 \|[\Ascr ^k u_0]_x\|_{PC^1} \leq c_3^{k+2}(k+2)! \FORALL k\geq1 \vspace{-1mm}
\end{equation}
and some $c_3>0$. Recall that $\theta^{-1},\sigma,\lambda\in PC^1[0,1]$. Using this, the estimates \eqref{eq:Aku01}-\eqref{eq:Aku02} and the fact $\|f g\|_{PC^1}\leq \|f\|_{PC^1} \|g\|_{PC^1}$ it follows from \eqref{eq:Akk+1} that: (i) $[\Ascr^k u_0]_{xx}\in PC^1[0,1]$ and $\|[\Ascr ^k u_0]_{xx}\|_{PC^1}\leq c_4^{k+2}(k+2)!$ for all $k\geq1$ and some $c_4>0$, (ii) the points where $[\Ascr^k u_0]_{xx}$ is discontinuous is a subset of $\Iscr$ and (iii) the points where $[\Ascr^k u_0]_{xx}$ is continuous but not differentiable (so that $[\Ascr^k u_0]_x$ is differentiable at these points) is also a subset of $\Iscr$. So there exists $c_5>0$ such that \vspace{-1mm}
\begin{equation} \label{eq:AkPC2est}
 \|\Ascr ^k u_0\|_{PC^{(2),1}}\leq c_5^{k+2}(k+2)! \FORALL k\geq1.
\end{equation}
Letting $\xi(x,t)=[\Ascr^k u_0](x)$ for $t\in[0,T]$ in \eqref{eq:pwiseJ1} and noting that $[B_n]_j=0$ for $j\in\{1,2,\ldots n-1\}$ gives \eqref{eq:AkAnknotJ}. \vspace{-1mm}
\end{proof}

The next proposition presents an estimate for the solution of a second-order difference equation that we encounter in this section as well as Section \ref{sec5}. \vspace{-1mm}

\begin{framed} \vspace{-2mm}
\begin{proposition}\label{prop:discretesystem}
Consider the discrete-time system \vspace{-2mm}
\begin{equation} \label{eq:DTSvar}
 \bbm{z_j \\ z_{j+1}} = \bbm{0 & 1 \\ -1+a_j h & 2-a_j h - b_j h^2 } \bbm{z_{j-1} \\ z_j} +\bbm{0 \\ \eta_j} \vspace{-2mm}
\end{equation}
for $j\geq2$ with initial state $\sbm{z_1 \\ z_2}=\sbm{p \\ p + q h}$. Here $p, q, a_j, b_j, \eta_j\in\rline$ and $h=1/(n+1)$ for some integer $n\geq 3$. There exists a $M$ which depends only on
$a=\sup_{2\leq j\leq n}|a_j|+1$ and $b=\sup_{2\leq j\leq n}|b_j|$ such that \vspace{-3mm}
\begin{equation}\label{eq:solzj}
  |z_j| \leq M \big(\,|p|+ j h |q| + \sum_{k=2}^{j-1} (j-k)|\eta_k|\,\big) \FORALL j\in\{3,4,\ldots n\}. \vspace{-6mm}
\end{equation}
\end{proposition}
\end{framed}
\vspace{-6mm}
\begin{proof}
Define the discrete-time system \vspace{-2mm}
\begin{equation} \label{eq:DTScon}
 \bbm{\zeta_j \\ \zeta_{j+1}} = \bbm{0 & 1 \\ -1-a h & 2+a h + b h^2} \bbm{\zeta_{j-1} \\ \zeta_j} + \bbm{0 \\ |\eta_j|} \qquad j\geq2, \vspace{-2mm}
\end{equation}
with initial state $\sbm{\zeta_1 \\ \zeta_2}=\sbm{|p| \\ |p| + |q|h}$.
From \eqref{eq:DTSvar} we have \vspace{-2mm}
\begin{equation} \label{eq:2DTS1}
 |z_{j+1}-z_j| = |(z_j-z_{j-1})(1-a_j h)-b_j h^2 z_j + \eta_j|. \vspace{-2mm}
\end{equation}
We will prove by mathematical induction that  \vspace{-1mm}
\begin{equation} \label{eq:majorMI}
 |z_j| \leq \zeta_j \quad {\rm and} \quad |z_j-z_{j-1}| \leq \zeta_j - \zeta_{j-1} \vspace{-1mm}
\end{equation}
for all $j\in\{2,3,\ldots n\}$. Clearly \eqref{eq:majorMI} holds for $j=2$. Suppose that \eqref{eq:majorMI} holds for all $j\in\{2,3,\ldots m\}$ for some $m<n$ (note that this implies $\zeta_1 \leq \zeta_2 \leq\cdots \zeta_m$). Then it follows from \eqref{eq:2DTS1} with $j=m$ and the definitions of $a$ and $b$ that \vspace{-1mm}
$$ |z_{m+1}-z_m| \leq (\zeta_m-\zeta_{m-1})(1+a h) + b h^2 \zeta_m + |\eta_m|. \vspace{-1mm}$$
From \eqref{eq:DTScon} with $j=m$, it is easy to verify that the right side of the above inequality is $\zeta_{m+1}-\zeta_m$, i.e. the second inequality in \eqref{eq:majorMI} holds for $j=m+1$. Using this we get $|z_{m+1}|\leq |z_{m+1}-z_m|+|z_m|\leq (\zeta_{m+1}-\zeta_m)+\zeta_m$, i.e. the first inequality in \eqref{eq:majorMI} also holds for $j=m+1$.
This completes the proof of \eqref{eq:majorMI} for $j\in\{2,3,\ldots n\}$.

Denote the state matrix in \eqref{eq:DTScon} by $S_h$. A simple calculation shows that $S_h$ has two real distinct eigenvalues $\lambda_{1h}=1+\alpha_{1h} h$ and $\lambda_{2h}=1-\alpha_{2h} h$ with corresponding eigenvectors $v_{1h}=\sbm{1 \\ \lambda_{1h}}$ and $v_{2h}=\sbm{1 \\ \lambda_{2h}}$, respectively, and $\alpha_{1h}$ and $\alpha_{2h}$ satisfy \vspace{-1mm}
\begin{equation} \label{eq:alphah}
 0 \leq \alpha_{1h},  \alpha_{2h} < \beta, \qquad \alpha_{1h}+\alpha_{2h} \geq 1. \vspace{-1mm}
\end{equation}
Here $\beta$ depends only on $a$ and $b$. Using this and writing $\sbm{r \\ s}\in \rline^2$ as a linear combination of $v_{1h}$ and $v_{2h}$ it follows that  \vspace{-2mm}
\begin{equation}\label{eq:merit}
  S_h^m\bbm{r \\ s} =  \frac{-\lambda_{2h} r + s}{\lambda_{1h}-\lambda_{2h}}\lambda_{1h}^m v_{1h} + \frac{\lambda_{1h} r -s}{\lambda_{1h}-\lambda_{2h}}\lambda_{2h}^m v_{2h} \vspace{-2mm}
\end{equation}
for any $m\geq0$. The solution to \eqref{eq:DTScon} is \vspace{-3mm}
\begin{equation} \label{eq:zetaconv}
 \bbm{\zeta_j \\ \zeta_{j+1}} = S_h^{j-1} \bbm{|p| \\ |p|+|q|h} + \sum_{k=2}^j S_h^{j-k} \bbm{0 \\ |\eta_k|} \qquad j\geq2. \vspace{-2mm}
\end{equation}
Note that $|\lambda_{1h}^m|, |\lambda_{2h}^m|<C$ for $0\leq m\leq n$, where $C$ depends only on $a$ and $b$. It follows via a simple calculation using \eqref{eq:alphah} and \eqref{eq:merit} that for $2\leq j\leq n$ the second element of $S_h^{j-1} \sbm{|p| \\ |p|}$ is bounded by $2C\beta|p|$ and the second element of $S_h^{j-1} \sbm{0\\ |q|h}$ is $|q|h \sum_{m=0}^{j-1} \lambda_{1h}^{j-1-m} \lambda_{2h}^m \leq |q|hj C^2$. Similarly for $2\leq k< j\leq n$ the second element of $S_h^{j-k} \sbm{0\\ |\eta_k|}$ is bounded by $|\eta_k|(j-k+1)C^2$.
Using these bounds in \eqref{eq:zetaconv} we get $\zeta_{j+1} \leq M[|p|+jh|q|+\sum_{k=2}^{j} |\eta_k|(j-k+1)]$ for $j\in\{2,3,\ldots n-1\}$ and some $M>0$ which, using \eqref{eq:majorMI}, implies the estimate in \eqref{eq:solzj}.
\end{proof}

The next theorem is the main result of this section. \vspace{-1mm}

\begin{framed} \vspace{-3mm}
\begin{theorem}\label{theorem:isa}
Let $u_0, u_T\in PC^{(2),1}(0,1)$ be as in the statement of Problem \ref{prob:motionplan} and let \eqref{eq:y_endpts1}-\eqref{eq:y_endpts2} hold for $y\in G_\alpha[0,T]$ with $1<\alpha<2$. Define $v_{n,1}\in G_\alpha[0,T]$ by taking $y_n=y$ in \eqref{eq:yn}. Then  $v_n(0)=[v_{n,1}(0) \ \ v_{n,2}(0) \ \cdots \ v_{n,n}(0)]^\top$ and  $v_n(T)=[v_{n,1}(T) \ \ v_{n,2}(T) \ \cdots \ v_{n,n}(T)]^\top$  determined for $n\gg1$ using $v_{n,1}$ and its derivatives and the expressions \eqref{eq:vn2flat}-\eqref{eq:vnjflat} satisfy the limits in \eqref{eq:vn0Tconv}. \vspace{-3mm}
\end{theorem}
\end{framed}

\begin{proof}
In this proof we suppose that $n\gg1$. Let $[z]_i$ denote the $i^{\rm th}$ component of $z\in\rline^n$. We will establish the first limit in \eqref{eq:vn0Tconv}; the second one can be proved analogously. Observe that the $n-1$ equations in \eqref{eq:vn2flat}-\eqref{eq:vnjflat} are precisely the first $n-1$ scalar equations in the vector ODE \eqref{eq:semi-disc}. From this, taking derivatives of \eqref{eq:semi-disc}, it is easy to verify that $v_n(0)$ defined in the theorem statement is the unique vector in $\rline^n$ such that $v_{n,1}^{(m)}(0)=[A_n^m v_n(0)]_1$ for $m\in\{0,1,\ldots n-1\}$. In our proof, we will take the vector $\xi_0=R_n u_0$ and modify its entries in $n$ steps so that the vector $\xi_p$ obtained after the $p^{\rm th}$-step satisfies $v_{n,1}^{(m)}(0)=[A_n^m \xi_p]_1$ for $m\in\{0,1,\ldots p-1\}$ and $[A_n^p \xi_p]_j=\Ascr^p u_0(j h)$ for $j\in\{1,2,\ldots n-p\}$, or equivalently $\xi_p$ satisfies \vspace{-1mm}
\begin{align}
 [A_n^m \xi_p]_1 &= \frac{\alpha_0\Ascr^m u_0(0) - \beta_0 (\Ascr^m u_0)_x(0)}{(\alpha_0 - \beta_0 q_0)} \FORALL m\in\{0,1,\ldots p-1\}, \label{eq:equmatch1} \\
 [A_n^p \xi_p]_j &=\Ascr^p u_0(j h) \FORALL j\in\{1,2,\ldots n-p\}, \label{eq:equmatch2} \\[-4.5ex]\nonumber
\end{align}
where $h=1/(n+1)$. We remark that the above equivalence follows using $y_n=y$, \eqref{eq:yn} and \eqref{eq:y_endpts1}. We will show that $\xi_p$ satisfies the estimate \vspace{-1.5mm}
\begin{equation} \label{eq:estkstep}
 \|R_n u_0 - \xi_p\|_{2d} \leq C h \vspace{-1.5mm}
\end{equation}
for some $C>0$ independent of $p$ and $n$. Clearly $\xi_n=v_n(0)$ and \eqref{eq:estkstep} with $p=n$ implies the first limit in \eqref{eq:vn0Tconv}. We will complete the proof in 3 steps.

\noindent
{\bf Step 1}. Given $\xi\in \rline^n$, $k\in\{1,2,\ldots n-1\}$ and constants $\{\rho_0, \rho_1, \ldots \rho_{n-k}\}$, we will show in this step that there exists a $\hat\xi \in\rline^n$ such that \vspace{-2mm}
\begin{align}
  [A_n^m \hat\xi]_1 & = [A_n^m \xi]_1 \FORALL m\in\{0,1,\ldots k-2\}, \label{eq:back1}\\
  [A_n^{k-1}\hat\xi]_1 &= [A_n^{k-1}\xi]_1+\rho_0, \label{eq:back2}\\
  [A_n^k \hat\xi]_j & = [A_n^k \xi]_j + \rho_j \FORALL j\in\{1,2,\ldots n-k\}. \label{eq:back3}\\[-4.3ex]\nonumber
\end{align}
Let $a_j = \sigma(jh)/\theta(jh)$ and $b_j =\lambda(jh)/\theta(jh)$ for $j\in \{2,3,\ldots n\}$. Suppose that for some $\tilde\xi\in \rline^n$ and each $p\in\{1,2,\ldots k\}$ the following set of equations hold: \vspace{-1mm}
\begin{align}
 [A_n^{k-p}\tilde\xi]_1 &\,= \rho_{0,p}, \label{eq:IC1}\\
 [A_n^{k-p}\tilde\xi]_2 &\,= \rho_{0,p} + q_0h\rho_{0,p} - h^2 \rho_{0,p} \frac{\sigma(h)q_0 + \lambda(h)}{\theta(h)(1-r_0)} + \frac{h^2\rho_{1,p}}{\theta(h)(1-r_0)}, \label{eq:IC2}
\end{align}
\vspace{-6mm}
\begin{align}
 \bbm{[A_n^{k-p}\tilde\xi]_j \\ [A_n^{k-p}\tilde\xi]_{j+1}}  &=  \bbm{0 & 1 \\ -1 + ha_j &2-ha_j - h^2b_j} \bbm{[A_n^{k-p}\tilde\xi]_{j-1} \\ [A_n^{k-p}\tilde\xi]_j} +\frac{h^2}{\theta(jh)} \bbm{ 0 \\  \rho_{j,p}} \nonumber \\
 &\hspace{35mm}  \FORALL j\in \{2,3,\ldots n-k+p-1\}, \label{eq:eODEk}\\[-4.4ex]\nonumber
\end{align}
where $\rho_{j,p}$ are defined as follows: $\rho_{j,1}= \rho_j$ for $j\in\{0,1,\ldots n-k\}$ and for $p>1$ we let $\rho_{0,p}=0$ and  $\rho_{j,p}=[A_n^{k-p+1}\tilde\xi]_j$ for $j\in\{1,2,\ldots n-k+p-1\}$. Then $\hat\xi=\tilde\xi+\xi$ satisfies \eqref{eq:back1}-\eqref{eq:back3}. Indeed, note that \eqref{eq:IC1} for $p\in\{1,2,\ldots k\}$ together with the definition of $\rho_{0,p}$ immediately implies that $\hat\xi$ satisfies \eqref{eq:back1}-\eqref{eq:back2}. For some $p\in\{1,2,\ldots k\}$ and each $j\in\{1,2,\ldots n-k+p-1\}$, write the equation $[A_n z]_j=\rho_{j,p}$ explicitly in terms of the parameters in \eqref{eq:heat1}-\eqref{eq:heat2} using the definition of $A_n$. Taking $z=A_n^{k-p}\tilde\xi$ and rearranging the terms in the resulting  equations we get the following $n-k+p-1$ equations: \eqref{eq:IC2} and the second row of \eqref{eq:eODEk} for $j\in\{2,3,\ldots n-k+p-1\}$. In other words for each $p\in\{1,2,\ldots k\}$, \eqref{eq:IC2} and \eqref{eq:eODEk} is equivalent to the equations $[A_n^{k-p+1} \tilde\xi]_j=\rho_{j,p}$ for $j\in\{1,2,\ldots n-k+1-p\}$. In particular, taking $p=1$, we get that $\tilde\xi$ satisfies $[A_n^{k} \tilde\xi]_j=\rho_j$ for $j\in\{1,2,\ldots n-k\}$, i.e. $\hat\xi=\tilde\xi+\xi$ satisfies \eqref{eq:back3}. So to find $\hat\xi$ which satisfies \eqref{eq:back1}-\eqref{eq:back3} it is sufficient to find $\tilde \xi$ which satisfies \eqref{eq:IC1}-\eqref{eq:eODEk} for $p\in\{1,2,\ldots k\}$.


Solve \eqref{eq:IC1}-\eqref{eq:eODEk} for $p\in\{1,2,\ldots k\}$
sequentially starting from $p=1$ and going to $p=k$ by ignoring the fact that $A_n^{k-p_1}\tilde\xi$ and $A_n^{k-p_2}\tilde\xi$ are related for different $p_1$ and $p_2$ (i.e. although the vectors $A_n^{k-p_1}\tilde\xi$ and $A_n^{k-p_2}\tilde\xi$ are both determined by $\tilde\xi$, we will treat them as independent vectors while solving  \eqref{eq:IC1}-\eqref{eq:eODEk} for $p=p_1$ and $p=p_2$). Denote the vector $A_n^{k-p}\tilde\xi$ obtained after solving  \eqref{eq:IC1}-\eqref{eq:eODEk} for any $p$ by $\tilde z^{k-p}$. We claim that $\tilde \xi=\tilde z^0$ satisfies \eqref{eq:IC1}-\eqref{eq:eODEk} for $p\in\{1,2,\ldots k\}$.
Indeed, $\tilde z^0$ clearly satisfies \eqref{eq:IC1}-\eqref{eq:eODEk} for $p=k$ with $\rho_{j,k}=\tilde z^1$. Suppose that $\tilde\xi=\tilde z^0$ satisfies \eqref{eq:IC1}-\eqref{eq:eODEk} for $p\in\{p^0,p^0+1,\ldots k\}$ for some $p^0>1$. Then using the argument in the previous paragraph (by taking $z=A_n^{k-p^0} \tilde z^0$) we get that $[A_n^{k-p^0+1} \tilde z^0]_j=\rho_{j,p^0}$ for $j\in\{1,2,\ldots n-k+1-p^0\}$, where $\rho_{j,p_0}= [\tilde z^{k-p_0+1}]_j$. Since $\tilde z^{k-p_0+1}$ is the solution of   \eqref{eq:IC1}-\eqref{eq:eODEk} for $p=p^0-1$ it follows that $\tilde z_0$ satisfies \eqref{eq:IC1}-\eqref{eq:eODEk} for $p=p^0-1$ also. This, via the principle of induction, completes the proof of our claim.

\noindent
{\bf Step 2}. Next we derive a bound for the $\tilde\xi$ satisfying \eqref{eq:IC1}-\eqref{eq:eODEk} for $p\in\{1,2,\ldots k\}$. 
Recall $\Iscr$ defined below \eqref{eq:heat2} and let $|\Iscr|$ be its cardinality. Given  $M_k>0$ such that $|\rho_0| \leq M_k h/\max\{1,|q_0|\}$ and all the $\rho_i$s in $\{\rho_1, \rho_2,\ldots \rho_{n-k}\}$, except less than $|\Iscr|$ of them, satisfy $|\rho_i|\leq M_k h$ and the rest of them satisfy $|\rho_i|\leq M_k$, we will show that \vspace{-1.5mm}
\begin{equation} \label{eq:backest}
 \|\tilde\xi\|_{2d} \leq \frac{L^k M_k h}{(2k-2)!}. \vspace{-3mm}
\end{equation}
Here $L>0$ depends only on the coefficients in \eqref{eq:heat1}-\eqref{eq:heat2}. For each $p\in \{1,2,\ldots k\}$, we regard \eqref{eq:eODEk} as a discrete-time system with initial conditions given in \eqref{eq:IC1}-\eqref{eq:IC2}. Applying Proposition \ref{prop:discretesystem} to this system, and using the expressions for $[A_n^{k-p}\tilde\xi]_1$  and $[A_n^{k-p}\tilde\xi]_2$ in \eqref{eq:IC1}-\eqref{eq:IC2}, we get that \vspace{-3mm}
\begin{equation}\label{eq:esttemp}
 |[A_n^{k-p}\tilde\xi]_j| \leq M \big(|\rho_{0,p}|+ |q_0\rho_{0,p}|+ jh^2|\rho_{1,p}|+h^2\sum_{i=2}^{j-1} (j-i)|\rho_{i,p}|\big) \vspace{-3mm}
\end{equation}
for all $j\in\{1,2,\ldots n-k+p\}$. Here $M$ depends only on the coefficients in \eqref{eq:heat1}-\eqref{eq:heat2} and we have used $jh\leq1$.

Let $p=1$. Using $\rho_{i,1}=\rho_i$ for $i\in\{0,1,\ldots n-k\}$ and the bounds on $\rho_i$ it follows from \eqref{eq:esttemp} that $|[A_n^{k-1} \tilde\xi]_j| \leq MM_k (3h+|\Iscr| h^2 n + h^3\sum_{i=2}^{j-1} (j-i))$, for all $j\in \{1,2,\ldots n-k+1\}$.
From this we get \vspace{-1.5mm}
\begin{equation} \label{eq:Ankbound}
 |[A_n^{k-1}\tilde\xi]_j| \leq 3N M_k h \quad \textrm{for} \quad  1\leq j \leq n-k+1.
 \vspace{-1.5mm}
\end{equation}
Here $N>M$ depends only on the coefficients in \eqref{eq:heat1}-\eqref{eq:heat2}. We will establish the following estimate for each $p\in \{2,3,\ldots k\}$: \vspace{-2mm}
\begin{equation}\label{eq:missing}
  [A_n^{k-p}\tilde\xi]_j = 0 \FORALL  1\leq j \leq p-1, \vspace{-1mm}
\end{equation}
\begin{equation}\label{eq:inductionest}
  |[A_n^{k-p}\tilde\xi]_j| \leq (3N)^p M_k \frac{j^{2p-2} h^{2p-1}}{(2p-2)!} \quad \forall\,  p\leq j \leq n-k+p. \vspace{-1mm}
\end{equation}
Let $p=2$. We have $\rho_{0,2}=0$ and $\rho_{i,2}=[A_n^{k-1} \tilde{\xi}]_i$ for $i\in \{1,2,\ldots n-k+1\}$. Using \eqref{eq:Ankbound} to bound $\rho_{i,2}$ it follows from \eqref{eq:esttemp} that \vspace{-4mm}
$$|[A_n^{k-2} \tilde\xi]_j|\leq 3NM_k M \Big(j h^3 + h^3 \sum_{i=2}^{j-1}(j-i)\Big) \vspace{-3mm}$$
for $j\in\{2,3,\ldots n-k+2\}$. From the above inequality, it follows that \eqref{eq:inductionest} holds for $p=2$. From \eqref{eq:IC1} we have $[A_n^{k-2}\tilde\xi]_1 =\rho_{0,2}$ and so \eqref{eq:missing} also holds for $p=2$.

Now suppose that \eqref{eq:missing}-\eqref{eq:inductionest} hold for $p \in \{2,3,\ldots \tilde{p}-1\}$ for some $2<\tilde p\leq k$. We have $\rho_{0,\tilde{p}}=0$ and $\rho_{i,\tilde{p}}=[A_n^{k-\tilde{p}+1}\tilde{\xi}]_i$ for $i\in \{1,2,\ldots n-k+\tilde{p}-1\}$. Using \eqref{eq:missing}-\eqref{eq:inductionest} with $p=\tilde{p}-1$  to bound $\rho_{i,\tilde p}$ (in particular we get $\rho_{1,\tilde{p}}=0$), it follows from \eqref{eq:esttemp} with $p=\tilde{p}$ that \vspace{-4mm}
$$|[A_n^{k-\tilde{p}} \tilde\xi]_j|\leq (3N)^{\tilde{p}-1} M M_k h^{2\tilde{p}-1} \sum_{i=2}^{j-1}(j-i) \frac{i^{2\tilde{p}-4}}{(2\tilde{p}-4)!} . \vspace{-3mm}$$
for $j\in\{\tilde p,\tilde p+1,\ldots n-k+\tilde p\}$.
For all integers $j\geq3$ and $1\leq p\leq j$ we have \vspace{-2.5mm}
\begin{equation} \label{eq:sumint}
 \sum_{i=2}^{j-1} (j-i) \frac{i^{2p-2}}{(2p-2)!} \leq \int_0^j (j+1-\tau) \frac{\tau^{2p-2}}{(2p-2)!}\dd\tau \leq 3\frac{j^{2p}}{(2p)!}. \vspace{-2.5mm}
\end{equation}
Using \eqref{eq:sumint} with $p=\tilde{p}-1$ to bound the summations in the estimate above \eqref{eq:sumint} we get that \eqref{eq:inductionest} holds for $p=\tilde{p}$. We have $\rho_{0,\tilde{p}}=0$ and \eqref{eq:missing} with $p=\tilde p-1$ gives $\rho_{i,\tilde{p}}=0$ for $i\in \{1,2,\ldots \tilde{p}-2\}$. Using this in \eqref{eq:IC1}-\eqref{eq:eODEk} we get that \eqref{eq:missing} also holds for $p=\tilde p$. So by the principle of induction it follows that \eqref{eq:missing}-\eqref{eq:inductionest} holds for all $p\in \{2,3,\ldots k\}$. From \eqref{eq:missing}-\eqref{eq:inductionest} with $p=k$ the estimate \eqref{eq:backest} follows.

\noindent
{\bf Step 3}. We will complete the proof of this theorem by showing that $\xi_p$ satisfying \eqref{eq:equmatch1}-\eqref{eq:estkstep} exists for $p\in\{1,2,\ldots n\}$. Recall \eqref{eq:AkPC2est}. Since $\Ascr^p u_0\in PC^{(2),1}[0,1]$, we get (i) $|\Ascr^p u_0(0) - \Ascr^p u_0(h)| \leq h\|\Ascr^p u_0\|_{PC^{(2),1}}$ and (ii) $|(\Ascr^p u_0)_x(0)-q_0\Ascr^p u_0(h)|\leq Kh \|\Ascr^p u_0\|_{PC^{(2),1}}$ for some $K>0$ ((ii) follows from \eqref{eq:nablan1} and \eqref{eq:appr_ux} with $u(x,t)=\Ascr^p u_0(x)$ and $j=1$). From (i)-(ii), the fact that $1/(\alpha_0 - \beta_0 q_0)$ and $q_0/(\alpha_0 - \beta_0 q_0)$ can be bounded by a constant independent of $n\gg1$, \eqref{eq:AkPC2est} and Lemma \ref{lm:lmestAku} we get \vspace{-1mm}
\begin{align}
 \Big|\frac{\alpha_0\Ascr^p u_0(0)-\beta_0 (\Ascr^p u_0)_x(0)}{\alpha_0 - \beta_0 q_0} - \Ascr^p u_0(h)\Big|&\leq \frac{h C_0^{p+2}(p+2)!} {\max\{1,|q_0|\}}, \label{eq:useest1} \\
 |\Ascr^{p+1} u_0(jh) - [A_n  R_n \Ascr^p u_0]_j| &\leq h^{ \delta_j} C_0^{p+2}(p+2)! \label{eq:useest2} \\[-4.3ex] \nonumber
\end{align}
for some $C_0>0$ and $j\in\{1,2,\ldots n-1\}$. Here $\delta_j$ is as in Lemma \ref{lm:lmestAku}. For each $p\geq0$, denote the term within the modulus sign in \eqref{eq:useest1} and \eqref{eq:useest2} by $\gamma_{0, p+1}$ and $\gamma_{j,p+1}$, respectively. Let $M_{p+1}=C_0^{p+2}(p+2)!$ and $C=\sum_{p=1}^\infty L^p M_p/(2p-2)!$.

From Step 1 there is a $\hat\xi\in\rline^n$ satisfying \eqref{eq:back2}-\eqref{eq:back3} with $k=1$, $\xi=\xi_0=R_n u_0$, $\rho_0=\gamma_{0,1}$ and $\rho_j=\gamma_{j,1}$. It is easy to check that  $\xi_1=\hat\xi$ satisfies \eqref{eq:equmatch1}-\eqref{eq:equmatch2} for $p=1$. Suppose a $\xi_q\in\rline^n$ satisfies \eqref{eq:equmatch1}-\eqref{eq:equmatch2} with $p=q$ for a  $q\in\{1,2,\ldots n-1\}$. From Step 1 there is a $\hat\xi\in\rline^n$  satisfying \eqref{eq:back1}-\eqref{eq:back3} with $k=q+1$, $\xi=\xi_q$,  $\rho_0=\gamma_{0,q+1}$ and $\rho_j=\gamma_{j,q+1}$. It is easy to verify that $\xi_{q+1}=\hat\xi$ satisfies \eqref{eq:equmatch1}-\eqref{eq:equmatch2} for $p=q+1$ and it follows from \eqref{eq:backest} that $\|\xi_{q+1}-\xi_q\|_{2d} \leq L^{q+1} M_{q+1} h/(2q)!$. So by the principle of induction there is a $\xi_p\in\rline^n$ satisfying \eqref{eq:equmatch1}-\eqref{eq:equmatch2} for each $p\in\{1,2,\ldots n\}$ and $\|\xi_p-\xi_{p-1}\|_{2d} \leq L^p M_p h/(2p-2)!$, which implies \eqref{eq:estkstep} with the $C$ given above. \vspace{-2mm}
\end{proof}
\vspace{-3mm}

\section{Motion planning} \label{sec5} \vspace{-1mm}
\setcounter{equation}{0} 

\ \ \ In our solution approach to the motion planning problem, Problem \ref{prob:motionplan}, presented in Section \ref{sec2.1} we have three claims: (i) there exists a $y\in G_\alpha[0,T]$ with $\alpha\in(1,2)$ satisfying \eqref{eq:y_endpts1}-\eqref{eq:y_endpts2}, (ii) the sequence of functions $\{f_n\}_{n=1}^\infty$ obtained by taking $y_n=y$ in \eqref{eq:param} converge to a $f\in C^1[0,T]$ and (iii) $f$ solves Problem \ref{prob:motionplan}. In this section, we will prove these claims in Propositions \ref{prop:yconst} and \ref{prop:anjconvj} and Theorem \ref{thm:finalmainresult}.

For $T\geq\Gamma>0$ and $\alpha\in(1,2)$, the following function  $\psi_0\in G_\alpha[0,T]$, see \cite{HuRo:1996}: \vspace{-1.5mm}
\begin{equation}\label{eq:psi0}
  \psi_0(t) =
  \begin{cases}
   0 & t\in [0,T]\setminus (0,\Gamma), \\
   \exp\left(-\left[\left(1-\frac{t}{\Gamma} \right)\frac{t}{\Gamma} \right]^{-\frac{1}{\alpha-1}} \right) & t\in (0,\Gamma).
  \end{cases} \vspace{-2mm}
\end{equation}

\begin{framed} \vspace{-2mm}
\begin{proposition}\label{prop:yconst}
Let $u_0$ and $u_T$ be as in Problem \ref{prob:motionplan}. For $\tau=0$ and $\tau=T$ define $y_{m,\tau}=\alpha_0 \Ascr^m u_\tau(0)-\beta_0(\Ascr^m u_\tau)_x(0)$ for each integer $m\geq0$. Then for $\tau\in\{0,T\}$, \vspace{-1mm}
\begin{equation} \label{eq:varC12}
  |y_{m,\tau}| \leq c_\tau^{m+2}(m+2)! \FORALL m\geq0
\end{equation}
and some $c_\tau>0$. Fix $\alpha\in(1,2)$ and $\Gamma\in(0,T]$ such that $\max\{c_0,c_T\}\Gamma<1$. Define $\psi_0$ via \eqref{eq:psi0} and let $\psi(t)=1-\big(\int_0^t \psi_0(\tau)\dd \tau\big) \big/ \big(\int_0^\Gamma \psi_0(\tau)\dd \tau\big)$ for $t\geq 0$. For $\tau\in\{0,T\}$ define $g_\tau:[0,T]\to\rline$ as follows: $g_\tau(t)=0$ for $t\in(\Gamma,T]$ and \vspace{-2mm}
\begin{equation} \label{eq:gtseries}
 g_\tau(t) = \sum_{m=0}^\infty y_{m,\tau} \frac{t^m}{m!}
\FORALL t\in[0,\Gamma]. \vspace{-2mm}
\end{equation}
Define $y(t)=g_0(t) \psi(t)+g_T(T-t) \psi(T-t)$ for $t\in[0,T]$. Then  $y\in G_\alpha[0,T]$ and it satisfies \eqref{eq:y_endpts1}-\eqref{eq:y_endpts2}. \vspace{-2mm}
\end{proposition}
\end{framed}
\vspace{-5mm}
\begin{proof}
The estimate \eqref{eq:varC12} for $\tau=0$ follows directly from \eqref{eq:AkPC2est}. Lemma \ref{lm:lmestAku} and its proof continue to hold if we replace $u_0$ with $u_T$ everywhere. In particular, \eqref{eq:AkPC2est} holds with $u_T$ in place of $u_0$ and that implies \eqref{eq:varC12} for $\tau=T$.

From $\psi_0\in G_\alpha[0,T]$ we get $\psi\in G_\alpha[0,T]$. 
Using \eqref{eq:varC12} we can check that the series in \eqref{eq:gtseries} is uniformly convergent on $[0,\Gamma]$, i.e. $g_0$ is a real analytic function on $[0,\Gamma]$, and so $g_0\in G_\alpha[0,\Gamma]$. Using $\psi\in G_\alpha[0,T]$, $g_0\in G_\alpha[0,\Gamma]$, $\psi^{(m)}(\Gamma)=0$ for $m\geq0$ (which is easy to check) and $g_0(t)=0$ for $t\in(\Gamma,T]$ we can conclude that $g_0\psi$ is infinitely differentiable on $[0,T]$. Combining this with the facts that $g_0\psi$ restricted to $[0,\Gamma]$ is in $ G_\alpha[0,\Gamma]$ (this is because $G_\alpha[0,\Gamma]$ is closed under multiplication) and $g_0\psi(t)=0$ for $t\in[\Gamma,T]$, we get that $g_0\psi\in G_\alpha[0,T]$. We can similarly show that $g_T\psi\in G_\alpha[0,T]$. Hence $y$ given by the expression below \eqref{eq:gtseries} is in  $G_\alpha[0,T]$. Differentiating this expression and using $\psi(0)=1$, $\psi(T)=0$,  $\psi^{(m)}(0)= \psi^{(m)}(T)=0$ for $m\geq1$ and \eqref{eq:gtseries}, it is easy to verify that $y$ satisfies \eqref{eq:y_endpts1}-\eqref{eq:y_endpts2}.
\end{proof}

Recall the expressions \eqref{eq:vn2flat}-\eqref{eq:fnflat}, which are equivalent to the semi-discrete system \eqref{eq:semi-disc}, and the coefficients $a_{n,k}$ from \eqref{eq:param}. For $k\leq1$, let $(2k-2)!=1$. \vspace{-1mm}

\begin{framed} \vspace{-2mm}
\begin{proposition}\label{prop:anjbounds}
There exists a $R>0$ independent of $k$ and $n$ such that for each $n\gg1$, \vspace{-1mm}
\begin{equation} \label{eq:anjbounds}
  |a_{n,k}| \leq \frac{R^{k+1}}{(2k-2)!} \FORALL k\in \{0,1,\ldots n\}.
  \vspace{-4mm}
\end{equation}
\end{proposition}
\end{framed}
\vspace{-6mm}
\begin{proof}
Fix $n\gg1$. From \eqref{eq:yn} we get $v_{n,1} = y_n/(\alpha_0-q_0\beta_0)$. From \eqref{eq:vn2flat}-\eqref{eq:fnflat} we get that $v_{n,j}$ for $j\in \{1,2,\ldots n\}$ and $f_n$ can be expressed as linear combinations of $v_{n,1}$ and its derivatives $v^{(1)}_{n,1}, v^{(2)}_{n,1},\ldots v^{(n)}_{n,1}$. Therefore there exist coefficients $d_{j,k}$ such that \vspace{-2mm}
\begin{equation}\label{eq:vnjparam}
  v_{n,j}=\sum_{k=0}^n d_{j,k}y_n^{(k)} \FORALL j\in \{1,2,\ldots n\} \vspace{-2mm}
\end{equation}
and \eqref{eq:param} holds. It is easy to check that \vspace{-1mm}
\begin{equation} \label{eq:djk0}
  d_{j,k} = 0 \quad {\rm if} \quad j\leq k. \vspace{-1mm}
\end{equation}
In \eqref{eq:fnflat}, replace $f_n$ with the summation from \eqref{eq:param} and replace $v_{n,n-1}$ and $v_{n,n}$ with the corresponding summation obtained from \eqref{eq:vnjparam}. Comparing the coefficient of $y_n^{(k)}$ on both the sides of the resulting expression gives \vspace{-1mm}
\begin{align*}
  a_{n,k} &= \frac{h^2}{b_n}d_{n,k-1} - \frac{(3r_1 - 1)\theta(nh)+ h^2\lambda(nh)}{b_n}d_{n,k}\\
  &\hspace{20mm} - \frac{h\sigma(nh) - (1-r_1)\theta(nh)}{b_n}(d_{n,k}-d_{n-1,k})\\[-4.5ex]
\end{align*}
for all $k\in \{0,1,\ldots n\}$. Here $d_{n,-1}=0$. A simple calculation using the  definitions of $b_n$ and $r_1$ presented above \eqref{eq:vn2flat} gives $1/b_n = (3\alpha_1+2h\beta_1)/(2h\theta(nh))$, $(1-r_1)/b_n = (\alpha_1 + h\beta_1)/(h\theta(nh))$ and $(3r_1-1)/b_n = -\beta_1/\theta(nh)$. Using these and the properties of the parameters in the parabolic PDE \eqref{eq:heat1}-\eqref{eq:heat2}, it follows from the above expression that there exists a constant $C_1>0$ independent of $n$ and $k$ such that \vspace{-2mm}
\begin{equation}\label{eq:ankest}
  |a_{n,k}| \leq C_1 \Big(h |d_{n,k-1}|  + |d_{n,k}| + \frac{|d_{n,k}-d_{n-1,k}|}{h}\Big) \vspace{-2mm}
\end{equation}
for all $k\in \{0,1,\ldots n\}$. We will complete the proof of this theorem by showing that there exist $R_1, R_2>0$ independent of $n$ and $k$ such that for each $k\in \{0,1,\ldots n\}$, \vspace{-2mm}
\begin{align}
  &|d_{j,k}| \leq \frac{R_1^{k+1}j^{2k}h^{2k}}{(2k)!} \FORALL j \in \{1,2, \ldots n\}, \label{eq:dnest} \\
  &|d_{j,k}-d_{j-1,k}| \leq \frac{hR_2^{k+1}}{(2k-2)!} \FORALL j \in \{2,3,\ldots n\}. \label{eq:ddnest}\\[-4.5ex]\nonumber
\end{align}
Note that bounding the terms on the right-side of \eqref{eq:ankest} using \eqref{eq:dnest}-\eqref{eq:ddnest} gives \eqref{eq:anjbounds}.

Using \eqref{eq:vnjparam} with $j=1$ in \eqref{eq:yn} we get $d_{1,0}= 1/(\alpha_0-q_0\beta_0)$. Replacing all the $v_{n,i}$ in \eqref{eq:vn2flat}-\eqref{eq:vnjflat} with the corresponding summation obtained from \eqref{eq:vnjparam} and comparing the coefficient of $y_n^{(k)}$ on both the sides of the resulting expressions gives \vspace{-1mm}
\begin{align}
  &d_{2,k} = \left(1 + q_0h - \frac{h^2[\sigma(h)q_0 + \lambda(h)]}{(1-r_0)\theta(h)}\right)d_{1,k}+\frac{h^2} {(1-r_0)\theta(h)}d_{1,k-1} , \label{eq:ICdk1}\\
  &d_{j+1,k} = \frac{-\theta(jh)+h\sigma(jh)}{\theta(jh)}d_{j-1,k} + \frac{2\theta(jh)-h\sigma(jh)-h^2\lambda(jh)}{\theta(jh)}d_{j,k} + \frac{h^2}{\theta(jh)}d_{j,k-1} \nonumber\\
   &\hspace{100mm} \forall\, j\in \{2,3,\ldots n-1\} \nonumber\\[-4.5ex] \nonumber
\end{align}
and for all $k\in \{0,1,\ldots n\}$. Here we take $d_{j,-1}=0$ for all $j$. For each $k\in \{0,1,\ldots n\}$, the above expressions can equivalently be written as \vspace{-2mm}
\begin{align}
  \bbm{d_{j,k} \\ d_{j+1,k}}  &=  \bbm{0 & 1 \\ -1 + ha_j &2-ha_j - h^2b_j} \bbm{d_{j-1,k} \\ d_{j,k}} + \bbm{0 \\ h^2d_{j,k-1}/\theta(jh)} \nonumber \\
  &\hspace{80mm} \forall \, j\in \{2,3,\ldots n-1\},\label{eq:dODEj} \\[-4.5ex] \nonumber
\end{align}
where $a_j = \sigma(jh)/\theta(jh)$ and $b_j = \lambda(jh)/\theta(jh)$. The initial state $[d_{1,k} \ \ d_{2,k}]^\top$ for this discrete-time system can be found from $d_{1,0}= 1/(\alpha_0-q_0\beta_0)$, \eqref{eq:djk0} and \eqref{eq:ICdk1}. Applying Proposition \ref{prop:discretesystem} to this system we get that there exists a $C_2>1$ that depends only on the coefficients in \eqref{eq:heat1}-\eqref{eq:heat2} such that for all $1\leq j \leq n$,  \vspace{-3mm}
\begin{equation}\label{eq:dj0est}
  |d_{j,k}| \leq C_2\big(|d_{1,k}|+ |q_0d_{1,k}| +j h^2|d_{1,k-1}| +  h^2 \sum_{i=2}^{j-1} (j-i)|d_{i,k-1}|\big).  \vspace{-3mm}
\end{equation}
Here we have used $jh\leq1$. \vspace{-1mm}

We will first show that \eqref{eq:dnest} holds for $k\in\{0,1,\ldots n\}$. Recall that $q_0=-\beta_0/(\alpha_0-\beta_0h)$. Substituting this in the expression for $d_{1,0}$ given above \eqref{eq:ICdk1} we get \vspace{-2mm}
$$ d_{1,0} = \frac{\alpha_0 - \beta_0h}{\alpha_0^2 + \beta_0^2 - \alpha_0\beta_0h}, \qquad q_0d_{1,0} = -\frac{\beta_0}{\alpha_0^2 + \beta_0^2 - \alpha_0\beta_0h}.  \vspace{-1mm} $$
From these expressions it follows that there exists a $C_3>1$ which depends only on the constants in \eqref{eq:heat2} such that $|d_{1,0}|\leq C_3$ and $|q_0d_{1,0}| \leq C_3$. Using this and $d_{j,-1}=0$ in \eqref{eq:dj0est} we get that \eqref{eq:dnest} holds for $k=0$ with $R_1=3 C_2 C_3$. Hence $|d_{j,0}| \leq R_1$ for $j\in\{1,2,\ldots n\}$. Using this and $d_{1,1}=0$ from \eqref{eq:djk0} in \eqref{eq:dj0est} gives $|d_{j,1}| \leq 3 C_2 R_1h^2j^2/2 \leq R_1^2 h^2j^2/2$, i.e. \eqref{eq:dnest} holds for $k=1$. Suppose that \eqref{eq:dnest} holds for $k=p-1$ with $2\leq p\leq n$. Then using \eqref{eq:dnest} with $k=p-1$ and $d_{1,p-1}=d_{1,p}=0$ (see \eqref{eq:djk0}) in \eqref{eq:dj0est} we get  \vspace{-2mm}
$$ |d_{j,p}| \leq C_2 R_1^{p}h^{2 p} \sum_{i=2}^{j-1} (j-i) \frac{i^{2p-2}}{(2p-2)!}.  \vspace{-2mm} $$
For $p \geq  j$, we have $d_{j,p}=0$ from \eqref{eq:djk0}. For $p < j$, using \eqref{eq:sumint} to bound the right side of the above inequality we get $|d_{j,p}|\leq 3 C_2 R_1^{p}h^{2p}j^{2p}/(2p)!$. Since $R_1>3 C_2$, it follows that \eqref{eq:dnest} holds for $k=p$. We can now conclude via the principle of mathematical induction that \eqref{eq:dnest} holds for each $k\in\{0,1,\ldots n\}$.

We will now complete the proof of this proposition by establishing \eqref{eq:ddnest}. Let $e_{j,k}=d_{j,k}-d_{j-1,k}$. For each $k\in \{0,1,\ldots n\}$, it follows from \eqref{eq:dODEj} that
$$ e_{j+1,k} = (1-h a_j)e_{j,k} - \eta_{j,k} \FORALL j\in \{2,3,\ldots n-1\},$$
where $\eta_{j,k} = h^2b_jd_{j,k} - h^2d_{j,k-1}/\theta(jh)$. Let $a=\max_{1\leq j\leq n}|a_j|$. Using the inequality $(1+a/n)^n \leq \exp(a)$, it is easy to see that the solution to the above first-order discrete-time system satisfies \vspace{-3mm}
\begin{equation} \label{eq:deldjk}
 |e_{j,k}| \leq \exp(a) \Big(|e_{2,k}| + \sum_{i=2}^{j-1} |\eta_{i,k}|\Big) \vspace{-3mm}
\end{equation}
for $2\leq j\leq n$. We get $|\eta_{j,k}|\leq C_4 h^2 R_1^{k+1}/(2k-2)!$ using \eqref{eq:dnest} and $|e_{2,k}| \leq C_5 h (|d_{1,k}|+ |q_0d_{1,k}| +h|d_{1,k-1}|)$ from \eqref{eq:ICdk1}, for some constants $C_4, C_5>0$ which depend only on the coefficients in \eqref{eq:heat1}-\eqref{eq:heat2}. It now follows, recalling $|d_{1,0}|<C_3$, $|q_0d_{1,0}|<C_3$, $d_{1,-1}=0$ and $d_{1,k}=0$ for $k\geq1$, that $|e_{2,k}| \leq 2C_3C_5 h/(2k-2)!$ for all $0\leq k\leq n$. Using these estimates for $\eta_{j,k}$ and $e_{2,k}$ in \eqref{eq:deldjk} immediately gives \eqref{eq:ddnest}.
\end{proof}

Taking $a_{n,k}=0$ for $k>n$ and letting $y_n=y$ in \eqref{eq:param} we get \vspace{-3mm}
\begin{equation}\label{eq:paramfull}
  f_n(t) = \sum_{k=0}^{\infty} a_{n,k} y^{(k)}(t). \vspace{-3mm}
\end{equation}
In the next result, using \eqref{eq:anjbounds}, we will prove that for any $y\in G_\alpha[0,T]$ with $1<\alpha<2$, the sequence of functions $\{f_n\}_{n=1}^\infty$ obtained from \eqref{eq:paramfull} converges in $C^1[0,T]$. \vspace{-1mm}

\begin{framed} \vspace{-2mm}
\begin{proposition} \label{prop:anjconvj}
Consider the coefficients $a_{n,k}$ introduced in \eqref{eq:paramfull}. For each $k\geq 0$ there exists an $a_k\in\rline$ such that \vspace{-1mm}
\begin{equation}\label{eq:lim_ank}
 \lim_{n\to \infty} a_{n,k} = a_k. \vspace{-1mm}
\end{equation}
Furthermore, for any $y\in G_\alpha[0,T]$ with $1<\alpha<2$, the sequence of functions $\{f_n\}_{n=1}^\infty$ obtained from \eqref{eq:paramfull} converges in $C^1[0,T]$ to $f=\sum_{k=0}^{\infty}a_k y^{(k)}$, i.e. \vspace{-1mm}
\begin{equation}\label{eq:limfn}
  \lim_{n\to\infty}\|f_n - f\|_{C^1[0,T]} = 0. \vspace{-6mm}
\end{equation}
\end{proposition}
\end{framed}
\vspace{-6mm}
\begin{proof}
From Proposition \ref{prop:anjbounds} we get that for each $k\geq0$ the sequence $\{a_{n,k}\}_{n=1}^\infty$ is bounded. So it has at least one converging subsequence. We will prove that all the converging subsequences of $\{a_{n,k}\}_{n=1}^\infty$ have the same limit, which will imply \eqref{eq:lim_ank}. To this end, for some $\tilde k\geq0$ we will suppose that $\{a_{n,\tilde k}\}_{n=1}^\infty$ has two converging subsequences with limits $a_{\tilde k}^1$ and $a_{\tilde k}^2$. It then follows using the Cantor diagonalization method that there exist two sequences $\{a^1_k\}_{k=0}^\infty$ and $\{a^2_k\}_{k=0}^\infty$ and two subsequences $\{n^1_p\}_{p=1}^\infty$ and $\{n^2_p\}_{p=1}^\infty$ of $\nline$ such that \vspace{-1.5mm}
\begin{equation}\label{eq:jmorek}
 \lim_{p\to\infty}a_{n^1_p,k} =a_k^1, \qquad \lim_{p\to\infty}a_{n^2_p,k}= a_k^2 \FORALL k\geq 0. \vspace{-1.5mm}
\end{equation}
We will prove \eqref{eq:lim_ank} by showing that $a^1_k=a^2_k$ for all $k\geq0$, in particular  $a^1_{\tilde k}=a^2_{\tilde k}$.

From \eqref{eq:anjbounds} and \eqref{eq:jmorek} we get $|a^1_k|, |a^2_k| \leq {R^{k+1}}/{(2k-2)!}$ for each $k\geq0$. Let $\psi$ be the function in Proposition \ref{prop:yconst} with $T=1$, $\Gamma=1$ and $\alpha=1.5$. For some integer $q\geq0$, let $y(t)= t^q \psi(1-t)$ for $t\in[0,1)$ and $y(t)=t^q\psi(t-1)$ for $t\in[1,2]$. Let $y(t)=0$ for $t>2$. Using the properties of $\psi$ (see proof of Proposition \ref{prop:yconst} for details) it is easy to differentiate $y$ and verify that $y\in G_{1.5}[0,T]$ for any $T>0$. Fix an integer $i\geq0$ and $T>0$. Using the bound on $a^1_k$ and the fact that $\|y^{(m)}\|_{C[0,T]}\leq D^{m+1} (m!)^{1.5}$ for some $D>0$ and all integers $m\geq0$, we get $\|a^1_k y^{(k+i)}\|_{C[0,T]} \leq M_k^i$ for each $k\geq0$ with $M_k^i=R^{k+1} D^{k+i+1}((k+i)!)^{1.5}/(2k-2)!$. Note that \vspace{-2mm}
$$ \lim_{k\to\infty}\frac{M_{k+1}^i}{M_k^i} = \lim_{k\to\infty} \frac{R D (k+i+1)^{1.5}}{(2k)(2k-1)}=0. \vspace{-2mm}$$
So it follows, via the ratio test, that $\sum_{k=0}^{\infty} M_k^i <\infty$. Now applying the Weierstrass $M$-test, we can conclude that the function series $\sum_{k=0}^{\infty} a^1_k y^{(k+i)}$ converges uniformly on $[0,T]$. We can similarly show that the function series $\sum_{k=0}^{\infty} a^2_k y^{(k+i)}$ converges uniformly on $[0,T]$.
Define \vspace{-3mm}
\begin{equation}\label{eq:f1f2}
  f^1(t) = \sum_{k=0}^{\infty} a^1_k y^{(k)}(t),  \qquad f^2(t) = \sum_{k=0}^{\infty} a^2_k y^{(k)}(t) \FORALL t\in[0,\infty). \vspace{-3mm}
\end{equation}
From the discussion above \eqref{eq:f1f2} we get that $f^1,f^2\in C^\infty[0,\infty)$, $f^{1(i)}=\sum_{k=0}^{\infty} a^1_k y^{(k+i)}$ and $f^{2(i)}=\sum_{k=0}^{\infty} a^2_k y^{(k+i)}$.
Define \vspace{-3mm}
\begin{equation} \label{eq:fnp1fnp2}
 f_{n_p^1}(t) = \sum_{k=0}^{n^1_p} a_{n^1_p,k} y^{(k)}(t), \qquad f_{n_p^2}(t) = \sum_{k=0}^{n^2_p} a_{n^2_p,k\,} y^{(k)}(t) \FORALL t\in[0,\infty). \vspace{-3mm}
\end{equation}
Let $T$ be as above. Fix $\epsilon>0$ and integer $i\geq0$. Choose $n_\epsilon$ such that $\sum_{k=n_\epsilon+1}^{\infty} M_k^i <\epsilon$ and then choose $p_\epsilon$ such that \eqref{eq:anjbounds} holds for $n\geq n_{p_\epsilon}^1$ and \vspace{-3mm}
$$\Big\|\sum_{k=0}^{n_\epsilon} (a_{n^1_p,k}-a_k^1) y^{(k+i)}\Big\|_{C[0,T]}<\epsilon \FORALL p>p_\epsilon. \vspace{-3mm}$$
Such a $n_\epsilon$ exists since $\sum_{k=0}^{\infty} M_k^i <\infty$ and such a $p_\epsilon$ exists since \eqref{eq:jmorek} holds. Then, since \eqref{eq:anjbounds} implies that $\|a_{n^1_p,k} y^{(k+i)}\|_{C[0,T]}\leq M_k^i$ for all $p\geq p_\epsilon$ and $\|a^1_k y^{(k+i)}\|_{C[0,T]} \leq M_k^i$ (shown above), it follows that \vspace{-3mm}
$$  \|f^{1(i)}-f^{(i)}_{n_p^1}\|_{C[0,T]} \leq \Big\|\sum_{k=0}^{n_\epsilon} (a_{n^1_p,k}-a_k^1) y^{(k+i)}\Big\|_{C[0,T]} + 2 \sum_{k=n_\epsilon+1}^\infty M_k^i < 3\epsilon \vspace{-3mm}$$
for all $p\geq p_\epsilon$. Since this is true for any $\epsilon>0$, we can conclude that $\lim_{p\to\infty} \|f^{1(i)}-f^{(i)}_{n_p^1} \|_{C[0,T]}=0$. Since $i\geq0$ can be any integer, we get \vspace{-3mm}
\begin{equation} \label{eq:f1tofnp}
 \lim_{p\to\infty}\|f^1-f_{n^p_1}\|_{C^m[0,T]}=0 \FORALL m\geq0. \vspace{-3.5mm}
\end{equation}
For all $m\geq0$, using the properties of $\psi$ we have $y^{(m)}(0)=0$ and so $f^{(m)}_{n_p^1}(0) =0$. Comparing the first expression in \eqref{eq:fnp1fnp2} with \eqref{eq:param}, it follows from the discussion below \eqref{eq:param} that the solution of \eqref{eq:semi-disc} with $n=n_p^1$, input $f_{n_p^1}$ and initial state $v_{n_p^1}(0)=0$ satisfies $v_{n_p^1,1}(t) = y(t)/(\alpha_0-q_0\beta_0)$ for $t\in[0,\infty)$. Therefore if $u^1$ is the solution of \eqref{eq:heat1}-\eqref{eq:heat2} with initial state $u_0=0$ and input $f=f^1$, it follows from the definition of $q_0$ and Corollary \ref{cor:semidisc_conv} that $u^1(0,t)=\alpha_0 y(t)/(\alpha_0^2+\beta_0^2)$ and $u_x^1(0,t)=-\beta_0 y(t)/(\alpha_0^2+\beta_0^2)$ for all $t\geq0$. We can similarly show that the solution $u^2$ of \eqref{eq:heat1}-\eqref{eq:heat2} with initial state $u_0=0$ and input $f=f^2$ satisfies $u^2(0,t)=\alpha_0 y(t)/(\alpha_0^2+\beta_0^2)$ and $u_x^2(0,t)=-\beta_0 y(t)/(\alpha_0^2+\beta_0^2)$ for all $t\geq0$.

Let $u$ be the solution of \eqref{eq:heat1}-\eqref{eq:heat2} with initial state $u_0=0$ and input $f=f^1-f^2$. Then $u(0,t)=u_x(0,t)=0$ for all $t\geq0$ and $f\in C^\infty[0,\infty)$ with
$f(t)=0$ for $t\geq2$. Let $\omega$ be the growth bound of the semigroup $\tline$ associated with \eqref{eq:heat1}. Taking the Laplace transform of \eqref{eq:heat1}, and denoting the transformed variable by a {\em hat}, we get \vspace{-1.5mm}
$$ s\hat{u}(x,s)  = \theta(x)\hat{u}_{xx}(x,s)+\sigma(x)\hat u_x(x,s) + \lambda(x)\hat{u}(x,s) \vspace{-1.5mm}$$
for all $s$ with $\Re s>\omega$ and almost all $x\in[0,1]$. From $u(0,t)=u_x(0,t)=0$ we get $\hat{u}(0,s)=\hat{u}_x(0,s) = 0$. Solving the above differential equation in $x$ gives $\hat u(x,s)=\hat u_x(x,s)=0$ for all $x\in[0,1]$ and $s$ with $\Re s>\omega$. In particular, $\alpha_1\hat u_x(1,s)+\beta_1\hat u(1,s)=\hat f(s)=0$ for all $s$ with $\Re s>\omega$. Hence $f(t)=0$ for all $t\geq0$. So $f(1)= \sum_{k=0}^q (a^1_k-a^2_k) q!/(q-k)!=0$. Since this expression holds for any $q$, it follows that $a^1_k=a^2_k$ for all $k\geq0$. This completes the proof of \eqref{eq:lim_ank}. \vspace{-1mm}

The limit \eqref{eq:limfn} can be shown my mimicking the steps used to prove \eqref{eq:f1tofnp}. \vspace{-2mm}
\end{proof}

We now present the main result of this paper. \vspace{-1.5mm}

\begin{framed} \vspace{-2mm}
\begin{theorem}\label{thm:finalmainresult}
Let $T$, $u_0$ and $u_T$ be as in Problem \ref{prob:motionplan}. Then there exists a $y\in G_\alpha[0,T]$ with $1<\alpha<2$ which satisfies \eqref{eq:y_endpts1}-\eqref{eq:y_endpts2}. Let the sequence of functions $\{f_n\}_{n=1}^\infty$ in $C^\infty[0,T]$ be obtained from \eqref{eq:paramfull}. Then there exists $f\in C^\infty[0,T]$ such that $\lim_{n\to\infty}\|f_n-f\|_{C^1[0,T]}=0$ and the state trajectory $u$ of \eqref{eq:heat1}-\eqref{eq:heat2} with $f$ as input and initial state $u(0)=u_0$ satisfies $u(T) = u_T$. \vspace{-2mm}
\end{theorem}
\end{framed}
\vspace{-6mm}
\begin{proof}
The existence of $y\in G_\alpha[0,T]$ with $1<\alpha<2$ which satisfies \eqref{eq:y_endpts1}-\eqref{eq:y_endpts2} follows from Proposition \ref{prop:yconst}. The existence of $f\in C^\infty[0,T]$ to which the sequence of functions $\{f_n\}_{n=1}^\infty$ converges in $C^1[0,T]$ follows from Proposition \ref{prop:anjconvj}.

Let $v_{n,1}(t) = y(t)/(\alpha_0-q_0\beta_0)$ for all $t\in [0,T]$. This, via \eqref{eq:vn2flat}-\eqref{eq:vnjflat}, determines a state trajectory $v_n$ of \eqref{eq:semi-disc} on the time-interval $[0,T]$   for the input $f_n$ given by \eqref{eq:paramfull} (or equivalently by \eqref{eq:param} with $y_n=y$). It now follows from Theorem \ref{theorem:isa} that $v_n(0)$ and $v_n(T)$ satisfy \eqref{eq:vn0Tconv}. Since $u_T\in C[0,1]$, we have $\lim_{n\to\infty}\|u_T-S_n R_n u_T\|_{L^2(0,1)}=0$. Also $\|z\|_{2d}=\|S_n z\|_{L^2(0,1)}$ for $z\in\rline^n$. Using these expressions along with the second limit in \eqref{eq:vn0Tconv} gives \vspace{-1mm}
\begin{equation}\label{eq:final1}
  \lim_{n\to\infty} \|u_T - S_n v_n(T)\|_{L^2}=\lim_{n\to\infty} \|R_n u_T - v_n(T)\|_{2d}=0. \vspace{-2mm}
\end{equation}
Let $\tilde v_n$ be the solution of \eqref{eq:semi-disc} with input $f_n$ and initial condition $R_n u_0$. Let $u$ be the state trajectory of \eqref{eq:heat1}-\eqref{eq:heat2} with $f$ as input and initial state $u(0)=u_0$. Applying Theorem \ref{th:semidisc_conv} we get \vspace{-1mm}
\begin{equation}\label{eq:final2}
 \lim_{n\to\infty} \|u(T) - S_n \tilde v_n(T)\|_{L^2}=0. \vspace{-1mm}
\end{equation}
Let $e_n(t)=\tilde v_n(t) - v_n(t)$. Then clearly $\dot e_n(t) = A_n e_n(t)$ with $e_n(0)=R_n u_0-v_n(0)$. From Proposition \ref{pr:eigenA} we get $\|e_n(T)\|_{2d}\leq C \|e_n(0)\|_{2d}$ for some constant $C$ independent of $n$. Combining this with the first limit in \eqref{eq:vn0Tconv}, while using $\|S_n z\|_{L^2(0,1)}=\|z\|_{2d}$ for $z\in\rline^n$, we get \vspace{-2mm}
\begin{equation}\label{eq:final3}
 \lim_{n\to\infty} \|S_n \tilde v_n(T) - S_n v_n(T)\|_{L^2}=0. \vspace{-2mm}
\end{equation}
Using the triangle inequality we have \vspace{-2mm}
\begin{align*}
 \|u(T) - u_T\|_{L^2} \leq &\,\|u(T) - S_n \tilde v_n (T) \|_{L^2} + \|S_n \tilde v_n(T) - S_n v_n(T) \|_{L^2} \nonumber \\
 &\quad  + \|S_n v_n(T) - u_T\|_{L^2}. \\[-4.6ex]
\end{align*}
By taking the limit as $n\to\infty$ on the right side of the above inequality and using \eqref{eq:final1}, \eqref{eq:final2}  and \eqref{eq:final3} we get $u(T)=u_T$. \vspace{-2mm}
\end{proof}

\section{Null controllability}\label{sec6}
\setcounter{equation}{0} 

\ \ \ The null controllability of the parabolic PDE \eqref{eq:heat1}-\eqref{eq:heat2} on any interval $[0,\tau]$ with $\tau>0$ follows directly from our solution to Problem \ref{prob:motionplan} presented in Theorem \ref{thm:finalmainresult}. Indeed, given $\tilde u_0\in L^2(0,1)$, fix $s\in(0,\tau)$ and let \eqref{eq:heat1}-\eqref{eq:heat2} evolve on the interval $[0,s]$ with initial state $u(0)=\tilde u_0$ and input $f=0$. Then the state of \eqref{eq:heat1}-\eqref{eq:heat2} at time $s$ is $\tline_s \tilde u_0$ which is in $\Dscr(\Ascr)$ and also in $PC^{(2),1}[0,1]$, see Lemma \ref{lm:reg}. Define $T=\tau-s$, $u_0=\tline_s \tilde u_0$ and $u_T=0$. Since $\tline$ is an analytic semigroup we have $u_0\in\Dscr(\Ascr^\infty)$, $\Ascr^ku_0 = \tline^{(k)}_s\tilde{u}_0$ for any integer $k>0$ and $u_0$ satisfies the estimates in \eqref{eq:probest}, see \cite[Chapter 2, Eq. (5.15)]{Pazy:1983}. Let $f\in C^\infty[0,T]$ be the input given by Theorem \ref{thm:finalmainresult} which transfers \eqref{eq:heat1}-\eqref{eq:heat2} from $u_0$ to $u_T$ over the time interval $[0,T]$. Then the state trajectory $\tilde u$ of \eqref{eq:heat1}-\eqref{eq:heat2} with initial state $\tilde u_0$ and input $\tilde f:[0,\tau]\to\rline$ defined as $\tilde f(t)=0$ for $t\in[0,s]$ and $\tilde f(t)=f(t-s)$ for $t\in(s,\tau]$ satisfies $\tilde u(\tau)=0$. Here the input $\tilde f\notin C^\infty[0,\tau]$ is a concatenation of two infinitely differentiable functions and the corresponding state trajectory $\tilde u$ is understood as the composition of the state trajectories corresponding to these inputs: i.e. $\tilde u(t)=\tline_t \tilde u_0$ for $t\in[0,s]$ and $\tilde u(t)=u(t-s)$ for $t\in[s,\tau]$, where $u$ is the state trajectory of \eqref{eq:heat1}-\eqref{eq:heat2} with initial state $\tilde u(s)$ and input $f$. We remark that null controllability of linear 1D parabolic PDEs with discontinuous coefficients has been proved in \cite{AleEsc:2008}, \cite{BeDeRo:2007}, \cite{FerZua:2002}, \cite{MaRoRo:2016}, \cite{Moy:2016} and \cite{Rous:2007}, often for coefficients less regular than those in this work, using approaches different from ours.

The construction of the input $f$ mentioned in the above paragraph requires the knowledge of the scalars $\Ascr^k \tline_s \tilde u_0(0)$ and $(\Ascr^k \tline_s \tilde u_0)_x(0)$ for all $k\geq0$. For $\tilde u_0\in PC[0,1]$ we will prove in Proposition \ref{prop:semireg} that
\vspace{-2mm}
\begin{align}
  \Ascr^k \tline_t \tilde u_0 (0) &= \lim_{n\to\infty} [A_n^ke^{A_n t}R_n \tilde u_0]_1, \label{eq:regular1}\\
  (\Ascr^k \tline_t \tilde u_0)_x(0) &= \lim_{n\to\infty} q_0[A_n^k e^{A_n t}R_n \tilde u_0]_1 \label{eq:regular2} \\[-4.6ex] \nonumber
\end{align}
for all $k\geq0$ and all $t\in [s,\tau]$. Here $A_n$ is the  matrix in \eqref{eq:semi-disc}, $q_0$ is defined above \eqref{eq:vn2flat} and the notations $R_n$ and $[\cdot]_1$ are as given at the end of Section \ref{sec1} and above Lemma \ref{lm:approxAq}, respectively. Inspired by \eqref{eq:regular1}-\eqref{eq:regular2} we will use $[A_n^ke^{A_n t}R_n \tilde u_0]_1$ and $q_0[A_n^k e^{A_n t}R_n \tilde u_0]_1$ instead of $\Ascr^k \tline_s \tilde u_0(0)$ and $(\Ascr^k \tline_s \tilde u_0)_x(0)$ to construct a sequence of functions, only by solving \eqref{eq:semi-disc}, such that their limit $\tilde g$ is an input which transfers \eqref{eq:heat1}-\eqref{eq:heat2} from $\tilde u_0\in PC[0,1]$ to $0$ over the time interval $[0,\tau]$, see Theorem \ref{th:nullcontrols}. (Recall that $PC[0,1]$ is the space of piecewise continuous functions.) In particular, for initial states in $PC[0,1]$ Theorem \ref{th:nullcontrols} provides a numerical scheme for obtaining approximations of the null control signal.



The next proposition establishes the uniform in $n$ analyticity of the semigroup $e^{A_n t}$. Let the induced norm of $A\in\rline^{n\times n}$ be $\|A\|_{2d}=\sup_{v\in\rline^n, \|v\|_{2d}=1}\|Av\|_{2d}$. \vspace{-2mm}

\begin{framed} \vspace{-2mm}
\begin{proposition}\label{prop:analytic}
The semigroup $e^{A_n t}$ generated by $A_n$ is analytic uniformly in $n$, i.e. for each $\tau>0$ there exists a constant $M_\tau>0$ such that \vspace{-2mm}
\begin{equation}\label{eq:analytic}
 \|A_n^k e^{A_n t}\|_{2d} \leq \frac{M_\tau^k k!}{t^k} \FORALL k\geq 0, \quad \forall\,n\gg1, \quad \forall\,t\in(0,\tau]. \vspace{-5mm}
\end{equation}
\end{proposition}
\end{framed}
\vspace{-6mm}
\begin{proof}
Recall that $A_n = \Theta_n L_n + \Sigma_n D_n + \Lambda_n$. Since the matrices $\Theta_n$, $\Sigma_n$, $\Lambda_n$ are diagonal and $L_n$ and $D_n$ are tridiagonal, it follows that $A_n$ is a tridiagonal matrix. Let $m_{i,j}$ denote $(i,j)$ entry of $A_n$.  Note that the immediate off diagonal entries of $A_n$ are given by $m_{1,2} = \theta(h)(1-r_0)/h^2$, $m_{i,i+1} = \theta(ih)/h^2$ and $m_{i,i-1} = (\theta(ih)-h\sigma(ih))/h^2$ for $i\in\{2,3,\ldots n-1\}$ and $m_{n,n-1}=(\theta(nh)(1-r_1) - h\sigma(nh))/h^2$. Define the diagonal matrix $P_n\in\rline^{n\times n}$ with diagonal entries $p_{1,1}=1$ and \vspace{-2mm}
\begin{equation} \label{eq:pii}
 p_{i,i}= \sqrt{\frac{m_{i,i-1}}{m_{1,2}}}\sqrt{\frac{m_{2,1}}{m_{2,3}} \frac{m_{3,2}}{m_{3,4}} \cdots \frac{m_{i-1,i-2}}{m_{i-1,i}}} \FORALL 2\leq i\leq n. \vspace{-2mm}
\end{equation}
A simple calculation shows that $P_n^{-1} A_n P_n$ is a symmetric matrix. (This is a standard approach to symmetrizing tridiagonal matrices, see for instance \cite[Chapter 3, Proposition 3.1]{GoMe:2010}.) Hence all the eigenvalues of $A_n$ are real and there exists an orthonormal matrix $Q_n$ such that $A_n = P_n Q_n E_n Q_n^\top P_n^{-1}$, where $E_n$ is a diagonal matrix containing all the eigenvalues $\{\lambda_1, \lambda_2, \ldots \lambda_n\}$ of $A_n$. For $n\gg1$, from the definitions of $r_0$ and $r_1$ given above \eqref{eq:vn2flat} we have $0.5\leq 1-r_0,1-r_1 \leq 1$. Using this, the facts that $\theta\in PC[0,1]$ is a positive function and $\sigma\in PC[0,1]$ it is easy to verify that for all $n\gg1$ there exist constants $c_1,c_2,c_3,c_4>0$ independent of $n$ such that $c_1<m_{i,i-1}/m_{1,2}<c_2$ for $i\in \{2,3,\dots n\}$ and $(1-c_3h)<m_{i,i-1}/m_{i,i+1}<(1+c_4h)$ for $i\in \{2,3,\ldots n-1\}$. Using these estimates, the expression $h=1/(n+1)$ and the bounds $(1-c_3 h)^n > e^{-c_3}$ and $(1+c_4 h)^n<e^{c_4}$ in \eqref{eq:pii} we get $c_1 e^{-c_3} < p_{i,i} < c_2 e^{c_4}$ for $i\in\{2,3,\ldots n\}$. Consequently, there exists $C>0$ independent of $n$ such that  $\|P_n\|_{2d}<C$ and $\|P_n^{-1}\|_{2d}<C$ for all $n\gg1$.

Observe that $A_n e^{A_n t} = P_n Q_n E_n e^{E_n t} Q_n^\top P_n^{-1}$ and $\|Q_n\|_{2d}=1$. So $\|A_n e^{A_n t}\|_{2d}\leq C^2 \|E_n e^{E_n t}\|_{2d} \leq C^2 \max \{|\lambda_i| e^{\lambda_i t}\m\big|\m i=1,2,\ldots n\}$. From Proposition \ref{pr:eigenA} we get that there exists an $\omega>0$ independent of $n$ such that $\lambda_i <\omega$  for $i\in\{1,2,\ldots n\}$. Hence  $\lambda_i e^{\lambda_i t} \leq \omega e^{\omega \tau}$ if $\lambda_i>0$ and $|\lambda_i| e^{\lambda_i t} \leq 1/t$ if $\lambda_i<0$ for $t\in(0,\tau]$ and so $\|A_n e^{A_n t}\|_{2d}\leq L_\tau/t$ for $t\in(0,T]$. Here $L_\tau=C^2 (1+\tau)\omega e^{\omega \tau}$. Since $A_n^k e^{A_n t}= (A_n e^{A_n t/k})^k$ due to the semigroup property, it follows that $\|A_n^ke^{A_nt}\|_{2d}\leq L_\tau^k k^k/t^k$. This and $k!e^k\geq k^k$ implies that \eqref{eq:analytic} holds with $M_\tau= L_\tau e$. \vspace{-2mm}
\end{proof}

We will now prove \eqref{eq:regular1}-\eqref{eq:regular2}. \vspace{-2mm}

\begin{framed} \vspace{-2mm}
\begin{proposition}\label{prop:semireg}
For each $\tilde u_0\in PC[0,1]$, $\tau>0$ and $s\in(0,\tau)$ the limits in \eqref{eq:regular1}-\eqref{eq:regular2} hold for all $k\geq0$ and $t\in[s,\tau]$ with the convergence being uniform in $t$. \vspace{-6mm}
\end{proposition}
\end{framed}
\vspace{-6mm}
\begin{proof}
For each $n\gg1$, $k\geq0$, $u_\infty \in \Dscr(\Ascr^\infty)$ and $t\in[0,\tau]$, define $e_{n,k}(u_\infty,t) = [R_n \Ascr^k\tline_t -A_n^k e^{A_n t}R_n]u_\infty$. Fix $s\in(0,\tau)$. We claim that \vspace{-2mm}
\begin{equation}\label{eq:limsmoothu}
  \lim_{n\to\infty} \sup_{t\in [s,\tau]}\|e_{n,k}(u_\infty, t)\|_{2d} = 0 \FORALL u_\infty \in \Dscr(\Ascr^\infty) \vspace{-2mm}
\end{equation}
and all $k\geq0$. Indeed, for any $u_\infty \in \Dscr(\Ascr^\infty)$ note that $e_{n,0}(u_\infty,t) = R_n \tline_t u_\infty-e^{A_n t}R_n u_\infty$ is precisely $e_n(t)$ defined below \eqref{eq:initcont} when the initial state and input for \eqref{eq:heat1}-\eqref{eq:heat2} are $u_0=u_\infty$ and $f=0$ and the initial state and input for \eqref{eq:semi-disc} are $R_n u_0=R_n  u_\infty$ and $f_n=0$ so that $u(t)=\tline_t u_\infty$ and  $v_n(t)=e^{A_n t}R_n u_\infty$. These initial states and input satisfy the conditions in Corollary \ref{cor:semidisc_conv} and so we can use \eqref{eq:en_rho0} to conclude that \eqref{eq:limsmoothu} holds for $k=0$. Now suppose that \eqref{eq:limsmoothu} holds for some $k_0\geq 0$. For any $u_\infty \in \Dscr(\Ascr^\infty)$, a simple calculation shows that \vspace{-1mm}
\begin{equation} \label{eq:enmind}
 e_{n,k_0+1}(u_\infty,t) = A_n^{k_0} e^{A_n t}(R_n \Ascr u_\infty - A_n R_n u_\infty) + e_{n,k_0}(\Ascr u_\infty,t). \vspace{-1mm}
\end{equation}
Since $u_\infty\in \Dscr(\Ascr^\infty)$, mimicking the steps above \eqref{eq:AkPC2est} with $k=0$ and $u_0=u_\infty$, we get $u_\infty\in PC^{(2),1}[0,1]$. Taking $\xi(\cdot,t)=u_\infty(\cdot)$ and $f_\xi=0$ in \eqref{eq:pwiseJ1one} we get $\lim_{n\to\infty}\|R_n \Ascr u_\infty - A_n R_n u_\infty\|_{2d}=0$ which, together with the estimate \eqref{eq:analytic}, implies that $\lim_{n\to\infty}\sup_{t\in[s,\tau]}\|A_n^{k_0} e^{A_n t}(R_n \Ascr u_\infty - A_n R_n u_\infty)\|_{2d}=0$. Using this and the expression $\lim_{n\to\infty}\| e_{n,k_0}(\Ascr u_\infty,t)\|_{2d}=0$ which follows from our assumption that \eqref{eq:limsmoothu} holds for $k_0$, we can conclude from \eqref{eq:enmind} that \eqref{eq:limsmoothu} holds for $k_0+1$. It now follows via the principle of induction that \eqref{eq:limsmoothu} holds for all $k\geq0$.

Given $\tilde u_0\in PC[0,1]$ and $s\in(0,\tau)$, define $e_{n,k}(\tilde u_0,t) = [R_n \Ascr^k\tline_t -A_n^k e^{A_n t}R_n]\tilde u_0$ for all $k\geq0$ and $t\in[s,\tau]$. We claim that \vspace{-2mm}
\begin{equation}\label{eq:limPCu}
  \lim_{n\to\infty} \sup_{t\in [s,\tau]}\|e_{n,k}(\tilde u_0, t)\|_{2d} = 0 \FORALL k\geq0. \vspace{-2mm}
\end{equation}
Indeed, for any fixed $k\geq0$ it follows from the analyticity of $\tline$ that there exists a constant $m_1>0$ such that $\sup_{t\in[s,\tau]}\|\Ascr^k\tline_t w\|_{\Dscr(\Ascr)} \leq m_1 \|w\|_{L^2}$ for all $w\in L^2(0,1)$. Since $\|w\|_{C[0,1]} \leq m_2 \|w\|_{H^2(0,1)} \leq m_3 \|w\|_{\Dscr(\Ascr)}$ for some constants $m_2, m_3>0$ and all $w\in\Dscr(\Ascr)$, we have $\sup_{t\in[s,\tau]} \|\Ascr^k\tline_t w\|_{C[0,1]} \leq m_k\|w\|_{L^2}$ for all $w\in L^2(0,1)$ with $m_k=m_1 m_2 m_3$. Fix $\epsilon>0$ and using the density of $\Dscr(\Ascr^\infty)$ in $L^2(0,1)$ choose $u_\infty\in\Dscr(\Ascr^\infty)$ such that $\|u_\infty-\tilde u_0\|_{L^2}<\epsilon$. Then clearly \vspace{-1mm}
\begin{equation} \label{eq:anaest1}
 \sup_{t\in[s,\tau]} \|R_n \Ascr^k\tline_t (u_\infty-\tilde u_0) \|_{2d} \leq m_k\epsilon. \vspace{-2mm}
\end{equation}
Since $u_\infty-\tilde u_0 \in PC[0,1]$, we have $\lim_{n\to\infty} \|R_n (u_\infty-\tilde u_0)\|_{2d}=\|u_\infty-\tilde u_0\|_{L^2[0,1]}$. It now follows from \eqref{eq:analytic} that there exists a constant $l_k>0$ such that \vspace{-2mm}
\begin{equation} \label{eq:anaest2}
 \limsup_{n\to\infty} \sup_{t\in[s,\tau]} \|A_n^k e^{A_n t} R_n (u_\infty-\tilde u_0)\|_{2d} \leq l_k\epsilon. \vspace{-1mm}
\end{equation}
Combining \eqref{eq:anaest1}-\eqref{eq:anaest2} with  \eqref{eq:limsmoothu} we get $\limsup_{n\to\infty} \sup_{t\in [s,\tau]}\|e_{n,k}(\tilde u_0, t)\|_{2d} \leq (m_k+l_k)\epsilon$. Since this estimate holds for all $\epsilon>0$, \eqref{eq:limPCu} follows.

Note that $A_n e_{n,k}(\tilde u_0,t) = e_{n,k+1}(\tilde u_0,t) - [R_n \Ascr^{k+1} \tline_t - A_n R_n \Ascr^k \tline_t] \tilde u_0$. Since  $\Ascr^k \tline_t \tilde u_0=\tline_{t-\frac{s}{2}} \Ascr^k \tline_{\frac{s}{2}} \tilde u_0$ for $t\in[s,\tau]$ and the analyticity of $\tline$ implies that $\Ascr^k \tline_{\frac{s}{2}} \tilde u_0 \in L^2(0,1)$, we conclude using Lemma \ref{lm:reg} that $\Ascr^k \tline_t \tilde u_0\in C([s,\tau];PC^{(2),1}[0,1])$. Furthermore, letting $u_0=\tline_t \tilde u_0$ in Lemma \ref{lm:lmestAku} it follows from the discussion above \eqref{eq:AkPC2est} that the points where $\Ascr^k \tline_t \tilde u_0$ is not differentiable is a subset of $\Iscr$ for all $t\in [s,\tau]$. Taking $\xi(\cdot,t)=\Ascr^k \tline_t \tilde u_0(\cdot)$ and $f_\xi=0$ in \eqref{eq:pwiseJ1one} we get $\lim_{n\to\infty}\sup_{t\in [s,\tau]}\|R_n \Ascr^{k+1} \tline_t \tilde u_0 - A_n R_n \Ascr^k \tline_t \tilde u_0\|_{2d}=0$ which, together with \eqref{eq:limPCu}, implies that \vspace{-0.5mm}
\begin{equation}\label{eq:AlimPCu}
  \lim_{n\to\infty} \sup_{t\in [s,\tau]}\|A_n e_{n,k}(\tilde u_0, t)\|_{2d} = 0 \FORALL k\geq0. \vspace{-1mm}
\end{equation}
Using \eqref{eq:limPCu} and \eqref{eq:AlimPCu}, it follows from \eqref{eq:estDen8} that \vspace{-0.5mm}
\begin{equation} \label{eq:enmdn}
 \lim_{n\to\infty}\sup_{t\in [s,\tau]}|[e_{n,k}(\tilde u_0,t)]_1|+|[D_n e_{n,k}(\tilde u_0,t)]_1| =0 \vspace{-1mm}
\end{equation}
for all $k\geq0$. Since $\Ascr^k \tline_t \tilde u_0\in C([s,\tau];PC^{(2),1}[0,1])$ we get \vspace{-0.5mm}
\begin{equation} \label{eq:amtcont}
 \lim_{n\to\infty}\sup_{t\in [s,\tau]}|\Ascr^k\tline_t \tilde u_0(0) - \Ascr^k\tline_t \tilde u_0 (h)| = 0 \vspace{-1mm}
\end{equation}
and, in addition using  \eqref{eq:nablan1} and \eqref{eq:appr_ux}, also that \vspace{-0.5mm}
\begin{equation} \label{eq:amtxcont}
 \lim_{n\to\infty}\sup_{t\in [s,\tau]}|(\Ascr^k\tline_t \tilde u_0)_x(0)-q_0\Ascr^k\tline_t\tilde u_0(h)| = 0. \vspace{-1.5mm}
\end{equation}
From the definitions of $e_{n,k}(\tilde u_0,t)$ and $D_n$ we have
$[e_{n,k}(\tilde u_0,t)]_1=\Ascr^k \tline_t \tilde u_0(h)-[A_n^k e^{A_nt}R_n \tilde u_0]_1$ and $[D_n e_{n,k}(\tilde u_0,t)]_1= q_0\Ascr^k \tline_t  \tilde u_0 (h)-q_0[A_n^k e^{A_nt}R_n \tilde u_0]_1$. It now follows using \eqref{eq:enmdn}, \eqref{eq:amtcont} and \eqref{eq:amtxcont} that \eqref{eq:regular1} and \eqref{eq:regular2} hold uniformly in $t\in [s,\tau]$.
\end{proof}

The next theorem presents a semi-discretization based numerical scheme for approximating the null control signals for \eqref{eq:heat1}-\eqref{eq:heat2} when its initial state belongs to  $PC[0,1]$. Recall the coefficients $a_{n,k}$ from \eqref{eq:paramfull} and $a_k$ defined in \eqref{eq:lim_ank}. \vspace{-1mm}

\begin{framed} \vspace{-2mm}
\begin{theorem} \label{th:nullcontrols}
Given $\tau>0$ and $\tilde u_0\in PC[0,1]$, fix $s\in (0,\tau)$ and let $T=\tau-s$.  Choose $\psi\in G_\alpha[0,T]$ with $1<\alpha<2$ such that  $\psi(0)= 1$, $\psi(T)=0$ and $\psi^{(k)}(0)=\psi^{(k)}(T)= 0$ for all $k\geq 1$. For each $n\geq1$, consider the following function which can be constructed by solving the semi-discrete system \eqref{eq:semi-disc}: \vspace{-3mm}
\begin{equation}\label{eq:gnyn}
  g_n(t) = \sum_{k=0}^{n}a_{n,k}y_n^{(k)}(t) \FORALL t\in [0,T], \vspace{-3mm}
\end{equation}
where $y_n(t) = (\alpha_0-\beta_0q_0)[e^{A_n(t+s)}R_n \tilde u_0]_1\psi(t)$. Then $g_n\in C^\infty[0,T]$ and there exists a function $g\in C^\infty[0,T]$ such that \vspace{-2mm}
\begin{equation}\label{eq:gnconv}
  \lim_{n\to\infty} \|g_n - g\|_{C^1[0,T]} =0. \vspace{-2mm}
\end{equation}
Define the function $\tilde g:[0,T]\to\rline$ by taking $\tilde g(t)=0$ for $t\in [0,s)$ and $\tilde g(t)=g(t-s)$ for $t\in [s,\tau]$. Then the state trajectory $\tilde u$ of \eqref{eq:heat1}-\eqref{eq:heat2} with initial state $\tilde u_0$ and input $\tilde g$ satisfies $\tilde u(\tau)=0$. \vspace{-2.5mm}
\end{theorem}
\end{framed}
\vspace{-2mm}
An example of a $\psi$ as in the theorem statement is presented in Proposition \ref{prop:yconst}. \vspace{-2mm}
\begin{proof}
Define $z_n(t)= e^{A_n(t+s)}R_n \tilde u_0$. From Proposition \ref{prop:analytic} it follows that there exists a constant $c_1>0$ such that $\|z_n^{(k)}(t)\|_{2d} \leq c_1^k k!$ and $\|A_n z_n^{(k)}(t)\|_{2d} \leq c_1^{k+1} (k+1)!$ for all $k\geq 0$, $n \gg 1$ and $t\in [0,T]$. Using these estimates in \eqref{eq:estDen8} (with $z_n^{(k)}$ in place of $e_n$) gives $|[z_n^{(k)}(t)]_1| \leq c_2^{k+1} (k+1)!$ and $|[D_n z_n^{(k)}(t)]_1| \leq c_2^{k+1} (k+1)!$ for some $c_2>0$ and all $k\geq 0$, $n \geq 1$ and $t\in [0,T]$. From the definition of $D_n$, which implies that $[D_n z_n^{(k)}(t)]_1= q_0[z_n^{(k)}(t)]_1$, and the inequality $(k+1)!\leq e^{k+1} k!$ it now follows \vspace{-1.5mm} that
\begin{equation} \label{eq:znkG}
 \sup_{t\in [0,T]} |(\alpha_0-q_0\beta_0)[z_n^{(k)}(t)]_1| \leq c^{k+1} k! \vspace{-2mm}
\end{equation}
for some $c>0$ and all $k\geq 0$ and $n \geq 1$. Hence $(\alpha_0-q_0\beta_0)[z_n]_1 \in G_1[0,T]$. Now since $\psi\in G_\alpha[0,T]$ for some $1<\alpha<2$ and $y_n=(\alpha_0-q_0\beta_0)[z_n]_1\psi$, using \cite[Proposition 1.4.5]{Rod:1993} and \eqref{eq:znkG}, it follows that $y_n\in G_\alpha[0,T]$ and there exists $L>0$ independent of $n$ such that \vspace{-1.5mm}
\begin{equation}\label{eq:ynest}
 \|y_n^{(k)}(t)\|_{C[0,T]} \leq L^{k+1} (k!)^\alpha \FORALL k\geq 0, \quad \forall\m n\geq 1. \vspace{-1.5mm}
\end{equation}
Define $\phi_n(t)=(\alpha_0-q_0\beta_0) [z_n(t)]_1$ and $\nu_k(t)=\alpha_0 \Ascr^k \tline_{t+s} \tilde u_0(0) - \beta_0(\Ascr^k \tline_{t+s} \tilde u_0)_x(0)$ for $t\in[0,T]$, $n\geq1$ and $k\geq0$.
Using Proposition \ref{prop:semireg} it follows that the sequence of ($k^{\rm th}$-derivative) functions $\{\phi^{(k)}_n\}_{n=1}^\infty$ converges to $\nu_k$ uniformly on $[0,T]$ for each $k\geq0$. Consequently we can conclude that (i) $\nu_0 \in C^\infty[0,T]$ with $\nu_0^{(k)}=\nu_k$ for each $k\geq0$, (ii) $y=\nu_0\psi\in C^\infty[0,T]$ and (iii) since $y_n=\phi_n\psi $, $y$ satisfies \vspace{-1.5mm}
\begin{equation}\label{eq:ynconv}
  \lim_{n\to\infty}\|y^{(k)}-y_n^{(k)}\|_{C[0,T]} =0 \FORALL k\geq 0. \vspace{-1.5mm}
\end{equation}
From \eqref{eq:ynest} and \eqref{eq:ynconv} we get \vspace{-1.5mm}
\begin{equation}\label{eq:yest_new}
 \|y^{(k)}(t)\|_{C[0,T]} \leq L^{k+1} (k!)^\alpha \FORALL k\geq 0 \vspace{-1.5mm}
\end{equation}
and so $y\in G_\alpha[0,1]$.

Let $R$ be as in \eqref{eq:anjbounds} and $L$ be as in \eqref{eq:ynest}. Define $M_{k,j} = \frac{R^{k+1}L^{k+j+1} ((k+j)!)^\alpha} {(2k-2)!}$. It follows from the ratio test that $\sum_{k=0}^{\infty}M_{k,j}<\infty$ for any $j\geq0$ (see discussion above \eqref{eq:f1f2}). From the estimates for $a_{n,k}$ and $a_k$ from \eqref{eq:anjbounds} and \eqref{eq:lim_ank} and $y_n^{(k)}$ and $y^{(k)}$ from \eqref{eq:ynest} and \eqref{eq:yest_new} we get that for all $k\geq0$, $j\geq0$ and $n\gg1$, \vspace{-1mm}
\begin{equation} \label{eq:heavyrain1}
 \|a_k y^{(k+j)}\|_{C[0,T]}\leq M_{k,j},
\end{equation}
\begin{equation} \label{eq:heavyrain2}
\|a_k y^{(k+j)}-a_{n,k} y_n^{(k+j)}\|_{C[0,T]}\leq 2M_{k,j}.
\end{equation}
Using $\sum_{k=0}^{\infty}M_{k,j}<\infty$ and \eqref{eq:heavyrain1}, it follows via a Weierstrass M-test that the series $\sum_{k=0}^{\infty} a_k y^{(k+j)}$ converges in $C[0,T]$ for all $j\geq0$. Hence the function $g=\sum_{k=0}^{\infty} a_k y^{(k)}$ belongs to $C^\infty[0,T]$. Fix $\epsilon>0$ and $k_\epsilon$ such that $\sum_{k=k_\epsilon+1}^{\infty} (M_{k,0}+M_{k,1})< \epsilon$. Then using \eqref{eq:gnyn} and \eqref{eq:heavyrain2} we get that \vspace{-3mm}
$$\limsup_{n\to\infty}\|g-g_n\|_{C^1[0,T]} \leq \limsup_{n\to\infty} \sum_{k=0}^{k_\epsilon}\|a_k y^{(k)}-a_{n,k} y_n^{(k)}\|_{C^1[0,T]}+2\epsilon=2\epsilon. \vspace{-3mm}$$
The last equality above follows from \eqref{eq:lim_ank} and \eqref{eq:ynconv}. So $\limsup_{n\to\infty}\|g-g_n\|_{C^1[0,T]}\leq2\epsilon$ for any $\epsilon>0$, i.e. \eqref{eq:gnconv} holds.

Define $u_0 = \tline_s \tilde u_0$ and $u_T=0$. From the analyticity of the semigroup $\tline$ it follows that $u_0 \in\Dscr(\Ascr^\infty)$ and satisfies the estimates in \eqref{eq:probest}, see \cite[Chapter 2, Eq. (5.15)]{Pazy:1983}. A simple computation shows that $y=\nu_0\psi$ defined above satisfies \eqref{eq:y_endpts1} and \eqref{eq:y_endpts2}. It now follows from Proposition \ref{prop:anjconvj} that the sequence of functions $\{f_n\}_{n=1}^\infty$ defined via \eqref{eq:paramfull} converge to $g$ in $C^1[0,T]$ and from Theorem \ref{thm:finalmainresult} that the state trajectory $u$ of \eqref{eq:heat1}-\eqref{eq:heat2} with input $g$ and initial state $u_0$ satisfies $u(T)=0$. Consequently the state trajectory $\tilde u$ of \eqref{eq:heat1}-\eqref{eq:heat2} with input $\tilde g$ defined in the theorem statement and initial state $\tilde u_0$ satisfies $\tilde u(\tau)=0$. \vspace{-2mm}
\end{proof}

\section{Numerical example} \label{sec7} \vspace{-1mm}
\setcounter{equation}{0} 

\ \ \ In the PDE \eqref{eq:heat1}-\eqref{eq:heat2} let $\theta(x)=1+x$ for $x\in[0,0.5)$, $\theta(x) = 2$ for $x\in [0.5,1]$, $\sigma(x)=\sin(5\pi x)$ for $x\in [0,0.3)$, $\sigma(x)=2-2x$ for $x\in [0.3,1]$, $\lambda(x)=e^{-5x}$ for $x\in [0,0.4)$, $\lambda(x)=2x^4$ for $x\in [0.4,1]$, $\alpha_0=\beta_1 =1$ and $\beta_0=\alpha_1=0$. Consider the problem of transferring \eqref{eq:heat1}-\eqref{eq:heat2} from the initial state $w_0$ defined as $w_0(x) = e^x\sin(2\pi x)$ for $x\in [0,0.3] \cup [0.7,1]$ and $w_0(x)=1-1/x$ for $x\in (0.3,0.7)$ to the steady state $w_\tau$ of \eqref{eq:heat1}-\eqref{eq:heat2} corresponding to the constant input $f_{ss}=0.5$ (see Remark \ref{rm:ss}) in time $\tau=0.5$ seconds. We compute $w_\tau$ numerically (Figure 2 at time $t=0.5$ shows $w_\tau$). We construct $f$ which transfers \eqref{eq:heat1}-\eqref{eq:heat2} from 0 to $w_\tau$ using Theorem \ref{thm:finalmainresult} and $\tilde g$ which transfers \eqref{eq:heat1}-\eqref{eq:heat2} from $w_0$ to $0$ using Theorem \ref{th:nullcontrols}. Clearly $r=f+\tilde g$ solves the problem. \vspace{-3mm}

$$\includegraphics[scale=1]{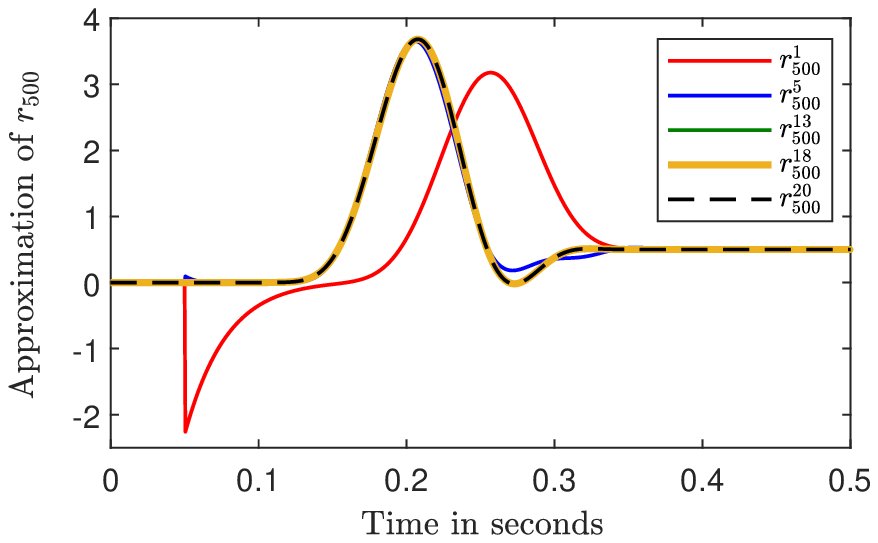}$$
\centerline{ \parbox{5.7in}{Figure 1. Plot of $r_{500}^i$ for $i\in \{1,5,13,18,20\}$. Here $\|r_{500}^{18}-r_{500}^{13}\|_{L^2[0,0.5]}<6.4\times 10^{-5}$ and $\|r_{500}^{20}-r_{500}^{18}\|_{L^2[0,0.5]}<1.4\times$\! $10^{-8}$. So a good approximation of $r_{500}$ is obtained by truncating the summation in \eqref{eq:paramfull} and \eqref{eq:gnyn} after a few terms. \vspace{1mm}}}

Let $n=500$ and find $d_{j,k}$ for $j\in\{2,3,\ldots 500\}$ by solving the difference equation \eqref{eq:dODEj}, starting with $k=0$ and going up to $k=500$, and then use the expression below \eqref{eq:djk0} to compute the coefficients $\{a_{500,k}\}_{k=0}^{500}$. In Theorem \ref{thm:finalmainresult} let $u_0 = 0$, $u_T = w_\tau$ and $y$ be the function obtained from Proposition \ref{prop:yconst} with $T= \Gamma = 0.5$ and $\alpha = 1.5$. Construct the function $f_{500}$ using \eqref{eq:paramfull}. In Theorem \ref{th:nullcontrols} let $s=0.05$, $\tilde u_0 = w_0$ and $\psi$ be the function in Proposition \ref{prop:yconst} with $T=0.45$ and $\alpha=1.5$. Construct $g_{500}$ using \eqref{eq:gnyn}. Define $\tilde g_{500}(t) = 0$ for $t\in[0,0.05)$ and $\tilde g_{500}(t) = g_{500}(t-0.05)$ for $t\in [0.05,0.5]$. Let $r_{500} = \tilde g_{500} + f_{500}$. Our choice $n=500$ is guided by our theoretical result that $r_n$ converges to $r$ as $n\to\infty$ and the numerical observation that $r_n$ for $n>500$ is very close to $r_{500}$. So $r_{500}$ is a good approximation of $r$.
The definition of $r_{500}$ contains 2 summations of $500$ terms each (from \eqref{eq:paramfull} and \eqref{eq:gnyn}). Let $r^i_{500}$ be the function obtained by truncating these summations after $i$ terms. Figure 1 shows that $r_{500}^{20}$ is a good approximation of $r_{500}$. The solution $u$ of \eqref{eq:heat1}-\eqref{eq:heat2} on the time interval $[0,0.5]$ with input $r_{500}^{20}$ and initial state $w_0$ is shown in Figure 2. The final state $u(0.5)$ is close to $w_T$ as expected from the Theorems \ref{thm:finalmainresult} and \ref{th:nullcontrols}. \vspace{-1mm}

$$\includegraphics[scale=1]{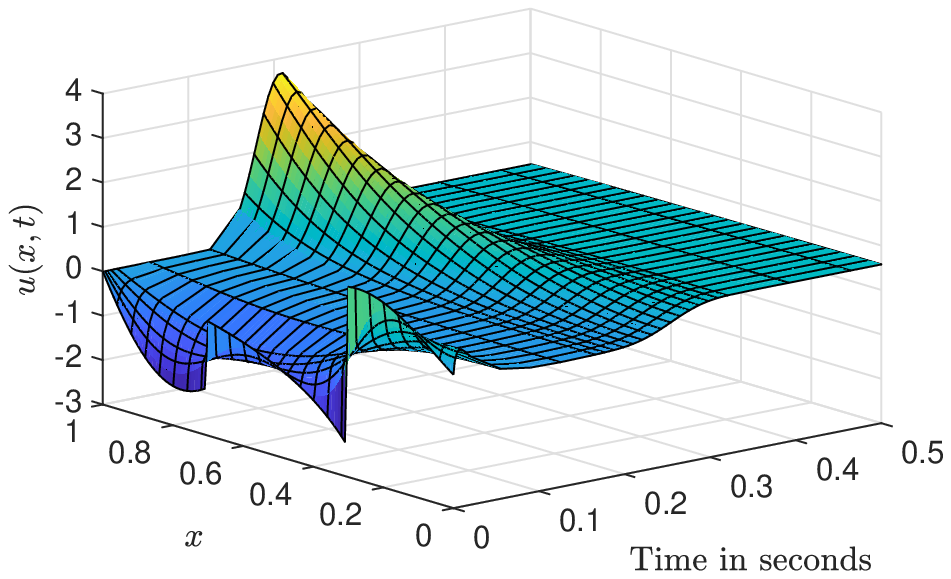}$$
\centerline{ \parbox{\textwidth}{\vspace{-1.5mm}
Figure 2. The solution $u$ of \eqref{eq:heat1}-\eqref{eq:heat2} with initial state $w_0$ and input $r_{500}^{20}$. The final state at $t=0.5$ seconds is close to $w_T$ ($\|u(0.5)-w_T\|_{L^2[0,1]} = 1.9\times 10^{-4}$). \vspace{-5mm}}}


{\small\begin{wrapfigure}[7]{L}{0.2\textwidth}
\centering\vspace{-4mm}
 \includegraphics[width=1in,height=1.25in,clip,keepaspectratio]{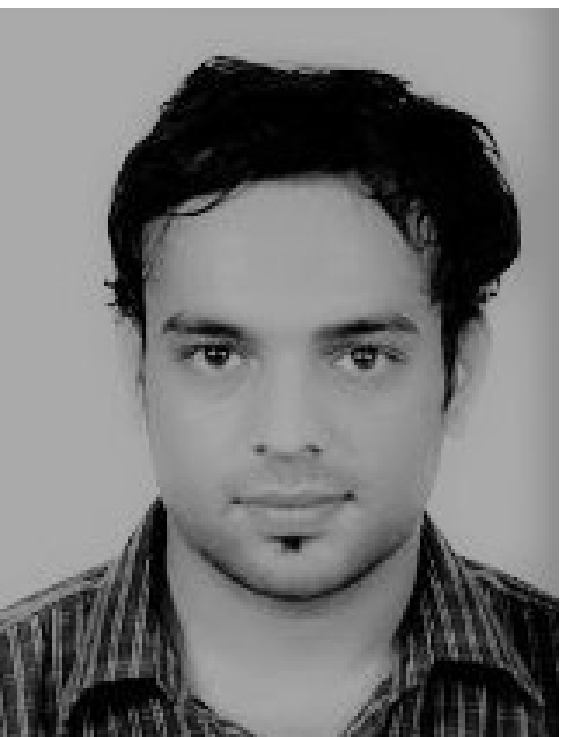}
\end{wrapfigure}

\noindent{\bf Soham Chatterjee} received the B.Tech. degree in instrumentation engineering from the West Bengal University of Technology in 2013, and the M.Tech. degree in electrical engineering from the Indian Institute of Technology Kanpur in 2016.

\noindent
He is currently pursuing the Ph.D. degree in Systems and Control Engineering at the Indian Institute of Technology Bombay. His research interests include control of distributed parameter systems. \vspace{2mm}

}

{\small \begin{wrapfigure}[8]{L}{0.2\textwidth}
\centering\vspace{-4mm}
\includegraphics[width=1in,height=1.25in,clip,keepaspectratio]{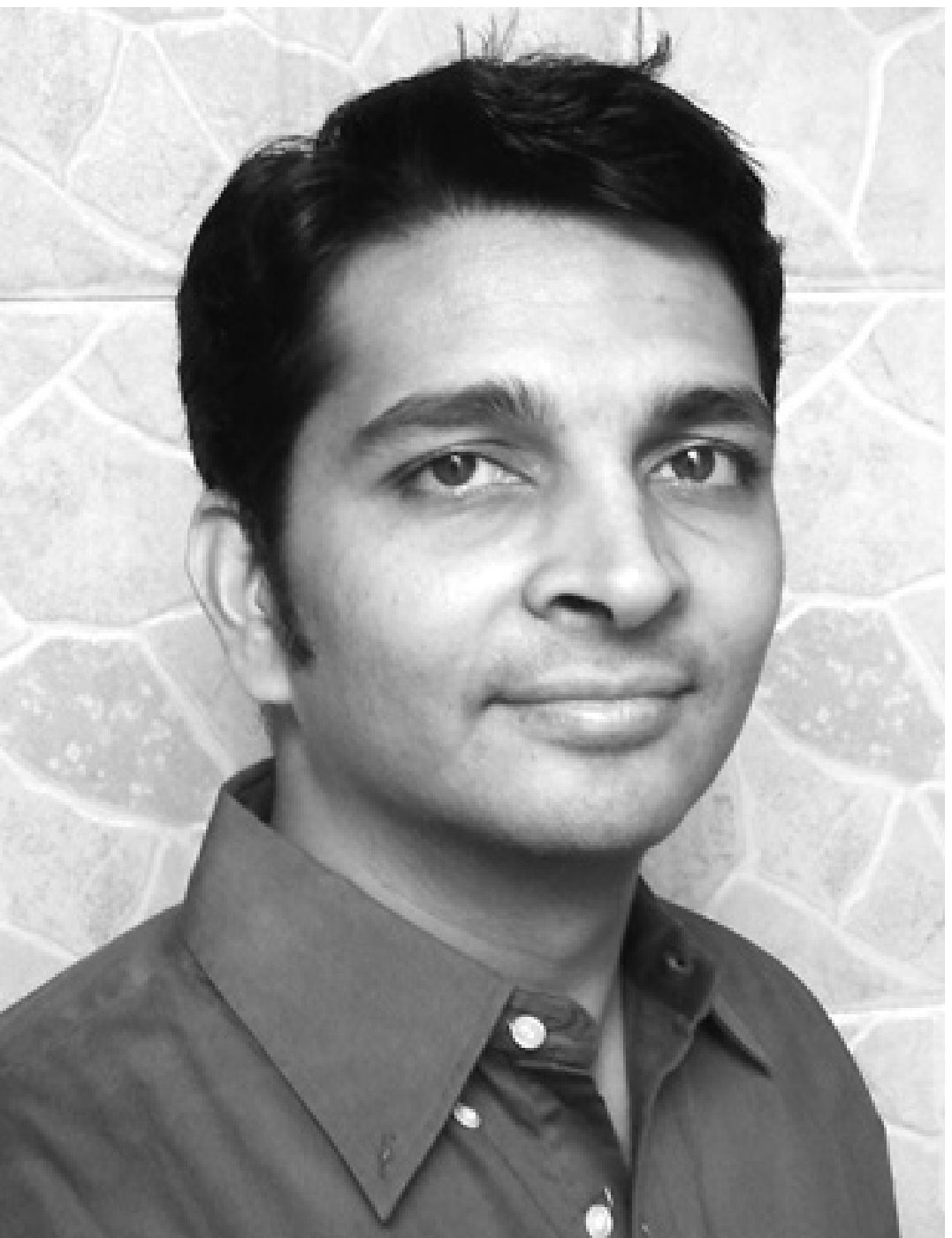}
\end{wrapfigure}

\noindent{\bf Vivek Natarajan} received the B.Tech. degree in mechanical engineering from the Indian Institute of Technology Madras in 2004, the M.S. degree in mechanical engineering from The Ohio State University in 2006, and the Ph.D. degree in mechanical engineering from the University of Illinois at Urbana-Champaign in 2012.

\noindent
He joined the Systems and Control Engineering group at Indian Institute of Technology Bombay in 2015, where he is currently an Associate Professor. Prior to that, he was a Postdoctoral Researcher in the School of Electrical Engineering, Tel Aviv University. His current research interests include control theory of infinite-dimensional systems.

}

\end{document}